\documentclass{aa}  

\usepackage{enumitem}  
\usepackage{titlesec}
\usepackage{multicol}
\usepackage{caption}
\usepackage{subfigure}
\usepackage{courier}
\usepackage{graphicx}
\usepackage{grffile} 
\usepackage{txfonts}
\usepackage{natbib}
\usepackage[flushleft]{threeparttable}

\titleformat{\subsection}{\bfseries\sffamily}{\thesubsection}{1em}{}

\begin{document}

   \title{Search and modelling of remnant radio galaxies in the \\ LOFAR Lockman Hole field}

   \author{M. Brienza\inst{1,2}\fnmsep\thanks{brienza@astron.nl}
          \and
          L. Godfrey\inst{1}
          \and
          R. Morganti\inst{1,2}
          \and 
          I. Prandoni\inst{3} 
          \and
          J. Harwood\inst{1}
          \and
          E. K. Mahony\inst{4,5} 
          \and
          M. J. Hardcastle\inst{6}
          \and 
          M. Murgia\inst{7}
          \and
          H. J. A. R\"ottgering\inst{8}
          \and
          T. W. Shimwell\inst{8}
          \and 
          A. Shulevski\inst{1}         
          }
          
   \institute{ASTRON, the Netherlands Institute for Radio Astronomy, Postbus 2, 7990 AA, Dwingeloo, The Netherlands
            \and
            Kapteyn Astronomical Institute, University of Groningen, PO Box 800, 9700 AV, Groningen, The Netherlands 
            \and
            INAF-ORA Bologna, Via P. Gobetti 101, 40129 Bologna, Italy
            \and
            Sydney Institute for Astronomy, School of Physics A28, The University of Sydney, NSW 2006, Australia
            \and
            ARC Centre of Excellence for All-Sky Astrophysics (CAASTRO), Australia
            \and
            Centre for Astrophysics Research, School of Physics, Astronomy and Mathematics, University of Hertfordshire, College Lane, Hatfield AL10 9AB, UK
            \and
            INAF-Osservatorio Astronomico di Cagliari, Loc. Poggio dei Pini, Strada 54, 09012 Capoterra (CA), Italy
            \and
            Leiden Observatory, Leiden University, P.O. Box 9513, 2300 RA Leiden, The Netherlands
            }

\abstract
{The phase of radio galaxy evolution after the jets have switched off, often referred to as the remnant phase, is poorly understood and very few sources in this phase are known.}
{In this work we present an extensive search for remnant radio galaxies in the Lockman Hole, a well-studied extragalactic field. 
We create mock catalogues of low-power radio galaxies based on Monte Carlo simulations to derive first-order predictions of the fraction of remnants in radio flux limited samples for comparison with our Lockman-Hole sample.}
{Our search for remnant radio galaxies is based on LOFAR observations at 150 MHz combined with public survey data at higher frequencies.
To enhance the selection process, and obtain a more complete picture of the remnant population, we use spectral criteria such as ultra-steep spectral index and high spectral curvature, and morphological criteria such as low radio core prominence and relaxed shapes to identify candidate remnant radio galaxies. Mock catalogues of radio galaxies are created based on existing spectral and dynamical evolution models combined with observed source properties.}
{We have identified 23 candidate remnant radio galaxies which cover a variety of morphologies and spectral characteristics. We suggest that these different properties are related to different stages of the remnant evolution. We find that ultra-steep spectrum remnants represent only a fraction of our remnant sample suggesting a very rapid luminosity evolution of the radio plasma. 
Results from mock catalogues demonstrate the importance of dynamical evolution in the remnant phase of low-power  radio galaxies to obtain fractions of remnant sources consistent with our observations. Moreover, these results confirm that ultra-steep spectrum remnants represent only a small subset of the entire population ($\sim$50\%) when frequencies higher than 1400 MHz are not included in the selection process, and that they are biased towards old ages.  }
{}


\maketitle


\section{Introduction}
\label{intro}

Radio-loud active galactic nuclei (AGN) are an episodic phenomenon in a galaxy's lifetime. The active phase of a radio AGN can last several tens of Myr, after which the radio jets stop and the source starts to fade away \citep{parma1999}. The fate of the radio galaxy remnant plasma and the physical processes driving its evolution are still poorly constrained, although they have implications for several areas of radio galaxy research. Firstly, the modelling of the radio spectrum of remnant sources provides constraints on the  timescales of activity and quiescence of the radio source, i.e. on its duty cycle, and on their dynamical evolution \citep{kardashev1962, murgia2011, kaiser2009, kapinska2015, turner2015}. Secondly, a better knowledge of the energetics of these objects can help quantify the role of radio AGN feedback, as well as give new insights into the formation of radio sources in galaxy clusters such as relics, halos, and phoenixes \citep{slee2001, ensslin2002, vanweeren2009, degasperin2015}. Larger samples of remnant radio galaxies are required to enable an investigation of their physical properties in a statistical sense and to provide new constraints on models describing the radio galaxy evolution.

Attempts have been made to find these sources  using all-sky surveys (e.g. \citealp{cohen2007}, \citealp{parma2007}, \citealp{murgia2011}) and individual deep fields (\citealp{sirothia2009}). 
Most of the searches have been based on spectral information. The radio spectrum of old remnant plasma is expected to be ultra-steep ($\rm  \alpha\gtrsim1.2 \ where \ S_\nu\propto\nu^{-\alpha} $) according to radiative cooling models (\citealp{pacholczyc1970}); therefore, ultra-steep spectral indices have been mostly used as the selection criterion (e.g. \citealp{cohen2007}, \citealp{parma2007}). \cite{murgia2011} suggested using the spectral curvature ($\rm SPC = \alpha_{high}-\alpha_{low}$) to select sources whose integrated spectrum is not yet ultra-steep below 1400 MHz but show a steepening at higher frequencies (\citealp{murgia2011}, \citealp{brienza2016}). This can happen if the period of time that has elapsed since the AGN switched off is much longer than the time the AGN was active. One of the main shortcomings of this method is the requirement for at least three different frequency observations at comparable resolution. 

However, recent results from mock radio catalogues derived from simulations of high-power radio galaxies by \cite{godfrey2017} show that spectral selection criteria only capture a fraction of the entire remnant population and are strongly biased towards very old sources. Therefore, complementary selection criteria should be considered  to create complete samples.

A few authors have based a search for remnant sources on morphological criteria alone without probing the spectral properties of the sources. For example, \cite{saripalli2012} selected sources that lack compact features like hot spots, jets, and cores, while \cite{giovannini1988} and \cite{hardcastle2016} use a criterion of low radio core prominence ($\rm S_{core}/S_{tot}<10^{-4}-5\times10^{-3}$). However, it is not yet clear whether these methods alone are able to select remnant sources efficiently.

In this paper we present a systematic search for remnant radio galaxies in one of the largest and best-characterized extragalactic deep fields, the Lockman Hole. This work makes use of the recent observations at 150 MHz performed with the LOw-Frequency ARray \citep[LOFAR,][]{vanhaarlem2013} published by \cite{mahony2016} (hereafter M16).
Thanks to its high sensitivity and excellent uv coverage, LOFAR is currently the best instrument for detecting sources with low surface brightness  at low frequency allowing us to characterize their morphology at high spatial resolution. 
Our motivation is to use this field to assess the selection strategy and to estimate how many remnants we will be able to discover in the LOFAR Two-metre Sky Survey (LoTSS, \citealp{shimwell2016}). 

In order to perform an extensive and ideally unbiased search for all classes of remnant radio galaxies, we adopt for the first time various selection criteria based  on spectral properties (ultra-steep spectral index and high spectral curvature) and on morphology (low radio core prominence and relaxed shapes). To do this we combine the LOFAR 150 MHz data with higher frequency public radio surveys, i.e. the 1400-MHz NRAO VLA Sky Survey, (NVSS, \citealp{condon1998}), the 325-MHz WENSS survey (\citealp{rengelink1997}), and the Faint Images of the Radio Sky at Twenty-cm survey (FIRST, \citealp{becker1995}).

In addition to the observational search, we also created mock catalogues of radio sources based on Monte Carlo  simulations to investigate how many remnant sources are expected in our flux limited sample. Following the work of \cite{godfrey2017} we created mock catalogues of low-power radio sources based on observed source properties and on published analytical radiative and dynamical evolution models of radio sources. The interesting aspect of this approach is that we can directly compare the empirical catalogue with the mock catalogue by applying the same flux density cut, and compare the results by applying the same selection criteria. 

The paper is organized as follows: in Section 2 we summarize the data used in this work; in Section 3 we describe the selection techniques and the results on the Lockman Hole; in Section 4 we present mock catalogues of low-power radio galaxies produced using Monte Carlo simulations to study the predicted fraction of remnant radio galaxies in the Lockman Hole. The cosmology adopted throughout the paper assumes a flat universe and the following parameters: $\rm H_{0}= 70\  km \ s^{-1}Mpc^{-1}$, $\Omega_{\Lambda}=0.7, \Omega_{M}=0.3$. 

\section{ Lockman Hole data}
\label{data}

Observations of the Lockman Hole field performed with the LOFAR High-Band Antennas (HBA) at 150 MHz have been recently published by M16. The sensitivity and resolution of these observations, as well as  the existence of ancillary data,  offer an excellent opportunity for our investigation on remnant radio galaxies. 
We summarize in Table \ref{tab:lock_mahony} the most relevant parameters of the observations and we refer the interested reader to M16 for a full description of the data reduction and the analysis of the field.

M16 presented the cross-match of the 150 MHz LOFAR catalogue with a deep (11-$\rm \mu Jy \ beam^{-1}$ rms) mosaic at 1400 MHz observed with the Westerbork Synthesis Radio Telescope (WSRT) and covering an area of 6.6 sq. degrees (\citealp{prandoni2017}). Hereafter, we will refer to this cross-matched catalogue as `Lockman-WSRT'. This allowed for a very sensitive spectral index study in the range 150-1400 MHz which will be used in our analysis. A spectral study of the sources with angular size < 40 arcsec was also performed on the entire LOFAR field of view (referred to as Lockman-wide) by combining the 150 MHz LOFAR catalogue with the NVSS, the WENSS, and the VLA Low-Frequency Sky Survey (VLSS, \citealp{cohen2007}).

\begin{table}[h]
        \centering
        \caption{Observation and image parameters and catalogue information for the Lockman Hole field from \cite{mahony2016}.  }
        \small
                \begin{tabular}{*2l}
                \hline
                \hline
                \textbf{LOFAR Observations}\\
                \hline
                RA (J2000) & 10h47m00s \\
                DEC (J2000) & +58$^o$05'00"\\ 
                Date of observation &18 March 2013\\
                Total observing time & 9.6 hrs \\
                Frequency range & 110-182 MHz\\
                \hline
                \textbf{LOFAR Image}\\
                \hline
                Field of view & 35 $\rm deg^2$\\
                Beam size & 18.6$\rm \times$14.7 arcsec, PA=85.7 deg \\
                Rms noise & 150-900 $\rm \mu$Jy$\rm\ beam^{-1}$\\
                Number of sources & 4882 \\
                \hline
                \textbf{WSRT Observations}\\
                \hline
                RA (J2000) & 10h52m16.6s \\
                DEC (J2000) & +58$^o$01'15"\\ 
                Date of observation & Dec 2006 - June 2007\\
                Total observing time & 12 hrs \\
                Frequency range & 1400 MHz\\
                \hline
                \textbf{WSRT Image}\\
                \hline
                Field of view & 6.6 $\rm deg^2$\\
                Beam size & 11$\rm \times$9 arcsec, PA=0 deg\\
                Rms noise & 11 $\rm \mu$Jy$\rm\ beam^{-1}$\\
                Number of sources & 6194 \\
                \hline
                \textbf{Lockman–WSRT catalogue}\\
                \hline
                Field of view & 6.6 $\rm deg^2$\\
                Number of LOFAR sources (150 MHz) & 1302\\
                Number of WSRT sources (1400 MHz) &1289\\
                \hline
                \hline  
                \end{tabular}
                \label{tab:lock_mahony}
\end{table}

\subsection*{New low-resolution catalogues}

In order to increase the image sensitivity to large-scale low surface brightness emission, which is typical of remnant radio galaxies, we have re-imaged the LOFAR data using the \texttt{awimager} software \citep{tasse2013}. We fix the pixel size to 8 arcsec, the weighting to robust=-0.5, and the longest baseline to 4 k$\rm \lambda$. Moreover, we apply a final restoring beam of 45 arcsec to match the resolution of NVSS. The rms at the centre of the field is 1.2 mJy $\rm beam^{-1}$ over a bandwidth of 50 MHz. With this new image we can more accurately cross-match all LOFAR sources with NVSS and WENSS to get spectral indices.
 
We perform the source extraction using the LOFAR Python Blob Detection and Source Measurement software (PYBDSM, \citealp{mohan2015}) following the same strategy used by M16. In this way we get a catalogue of 2588 sources above 5$\sigma$. To check the flux scale of the extracted catalogue we cross-match it with the GMRT 150 MHz All-sky Radio Survey alternative data release (TGSS ADR1, \citealp{intema2017}) using only sources that are point-like in the TGSS ADR1. We find an average flux density excess in the LOFAR sources of 7\% in agreement with M16. We therefore correct the final total flux densities of our catalogue by this factor. The uncertainties of the LOFAR flux densities are computed following M16.

To perform the spectral analysis we cross-match the 45 arcsec resolution LOFAR catalogue with higher frequency surveys using the Tool for OPerations on Catalogues And Tables (TOPCAT, \citealp{topcat}). Our spectral index analysis is based on two catalogues which are described below and summarized in Table~\ref{tab:lock_this}.

\medskip
The first catalogue (hereafter L45N) was created to investigate the spectral index distribution $\rm \alpha_{150}^{1400}$ of the sources in the entire LOFAR field, including extended sources, and to identify ultra-steep spectrum sources on a broad frequency range. 
The LOFAR catalogue at 45 arcsec resolution has been cross-matched with the NVSS catalogue using a search radius of 15 arcsec (M16). In order to account for the shallower flux limit of the NVSS, we only included sources with $\rm S_{peak,150}$>40 mJy so that all LOFAR sources without an NVSS counterpart (at 5$\rm \sigma$=2.3 mJy) have spectra steeper than $\rm \alpha_{150}^{1400}$=1.2. By making this cut the LOFAR catalogue is restricted to 543 sources. All sources have been visually inspected to check for any misidentifications in the automatic matching procedure and to make sure that multi-component sources are compared consistently between the two catalogues. As a result of this procedure we obtain an NVSS counterpart for 534 out of 543 sources.

Two-point spectral indices are calculated using the total flux density. For the nine objects that do not have a high-frequency counterpart we place an upper limit at 1400 MHz equal to 2.3 mJy ($\rm 5\sigma$ in NVSS). For these objects the spectral index is computed using the peak flux density at both frequencies. The error on the spectral index is computed using 

\begin{equation}
\label{erralpha}
\rm \alpha_{err} = \frac{1}{ln\frac{\nu_1}{\nu_2}}\sqrt{\left(\frac{S_{1,err}}{S_1}\right)^2+\left(\frac{S_{2,err}}{S_2}\right)^2}
,\end{equation}

where $S_1$ and $S_2$ are the flux densities at frequencies $\nu_1$ and $\nu_2$, and $S_{1,err}$ and  $S_{2,err}$ are the respective errors.

\begin{figure}
\centering
{\includegraphics[width=9cm]{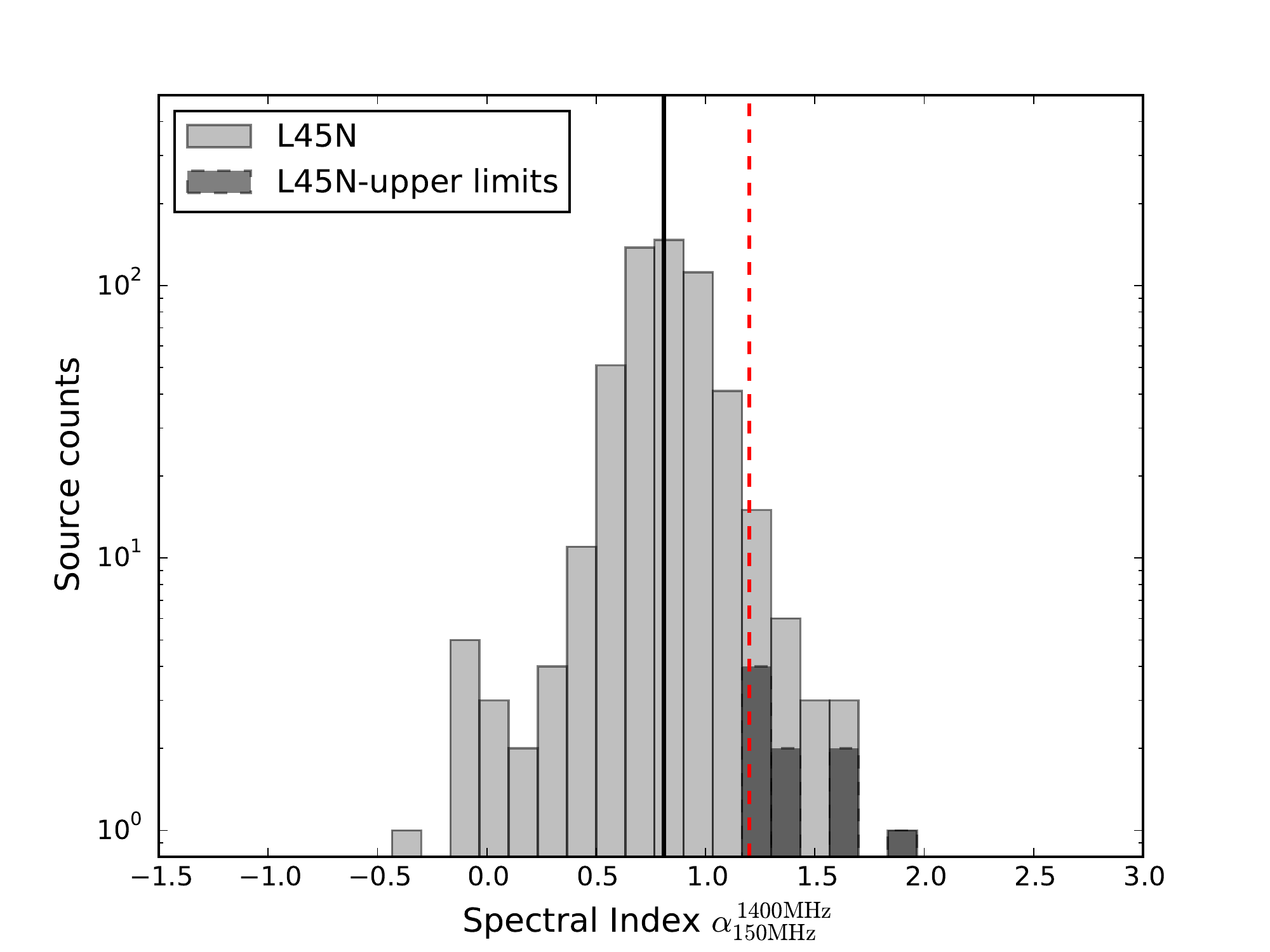}}
\caption{Spectral index distribution $\rm \alpha^{1400}_{150}$ calculated between LOFAR and NVSS using the L45N catalogue. Spectral indices of sources without a detection at 1400 MHz are here included as upper limits and are shown in black. The solid line represents the median value of the distribution equals to 0.81. The dashed line represents the boundary between steep and ultra-steep sources equal to $\rm \alpha^{1400}_{150}=1.2$ }
\label{fig:isto_alpha}
 \end{figure}

The spectral index median value is $\rm \alpha^{1400}_{150}=0.81\pm0.012$ (errors from bootstrap) with an interquartile range of [0.79, 0.83] in agreement with M16 (Figure \ref{fig:isto_alpha}).

\medskip
The second catalogue (hereafter L45NW) was created to investigate the spectral curvature of sources in the LOFAR field, including extended sources. To produce it the catalogue L45N is further cross-matched with the WENSS catalogue using a search radius of 15 arcsec (M16). Again, all sources have been visually inspected to check for any misidentifications in the automatic matching procedure and to make sure that multi-component sources are compared consistently between the catalogues. Out of 543 sources, 452 are found to have a WENSS counterpart. When using this catalogue in the following analysis we neglect sources that do not have a detection at 325 MHz. These sources would have both $\rm \alpha^{1400}_{325}$ and $\rm \alpha^{325}_{150}$ unconstrained, making it difficult to compute a useful value of spectral curvature. Sources only missing the NVSS counterpart are instead kept in the sample. For these sources only the spectral index $\rm \alpha^{1400}_{325}$ is a lower limit and the spectral curvature is consequently a lower limit as well. This allows us to include sources that have a sudden steepening in the spectrum (e.g. \citealp{brienza2016}). Two-point spectral indices, $\alpha^{1400}_{325}$ and $\alpha^{325}_{150}$, are calculated using the total flux density for the LOFAR and NVSS catalogue. For the WENSS catalogue we use the peak flux for all sources with major axis < 65 arcsec (the WENSS beam size in the Lockman Hole direction) and the total flux otherwise. Because of the lower sensitivity of the WENSS catalogue, we have verified that the peak flux density better represents the real flux densities of the sources especially at low flux levels. This is further justified by the lower resolution with respect to NVSS and LOFAR. We note that \cite{scaife2012} suggest that   the WENSS flux should be scaled by a factor of 0.9 (average on the entire sky) to match the LOFAR flux scale of \cite{roger1973}. Following M16, we do not to apply this correction because it causes a systematic underestimation of the WENSS flux density with respect to the value expected from the spectral index $\rm \alpha^{1400}_{150}$ for a typical radio source. The error on the spectral index is computed using Eq.~\ref{erralpha}.

\begin{table}[t]
        \centering
        \caption{Image parameters and catalogue information for the Lockman Hole field created for this work. }
        \small
                \begin{tabular}{*2l}
                \hline
                \hline
                \textbf{Image}\\
                \hline
                Beam size & 45$\rm \times$45 arcsec\\
                Rms noise & 1.2  mJy $\rm beam^{-1}$\\
                Number of sources & 2588 \\
                \hline
                \textbf{L45N catalogue}\\
                \hline
                Field of view & 35 $\rm deg^2$\\
                Number of LOFAR sources (150 MHz) & 543\\
                \tiny
                $\rm (Sources \ with \ S_{peak}>40 \ mJy)$ &\\
                Number of NVSS sources (1400 MHz) &534\\
                \hline
                \textbf{L45NW catalogue}\\
                \hline
                Field of view & 35 $\rm deg^2$\\
                Number of LOFAR sources (150 MHz)&452 \\
                \tiny
                $\rm (Sources \ with \ S_{peak}>40 \ mJy)$ &\\          
                Number of WENSS sources (325 MHz)&452\\
                Number of NVSS sources (1400 MHz) &444\\
                \hline
                \hline  
                \end{tabular}
                \label{tab:lock_this}
\end{table}

\section{Selection of remnant radio galaxies}
\label{selection}

In this section we describe the approach that we used to select remnant radio galaxies in the Lockman Hole field. The identification of this class of sources is challenging, due to the variety of characteristics that they are expected to have, which depend on their age and physical properties.

\cite{godfrey2017} has demonstrated that the spectral selection criteria are biased towards old remnants and do not allow us to select the entire remnant population, especially without frequencies higher than 5000 MHz where the steepening occurs sooner. For this reason we use here  spectral criteria such as ultra-steep spectral index and high spectral curvature, and morphological criteria such as low radio core prominence (CP) and relaxed shapes to identify remnant candidates. The combination of remnants identified by these techniques can provide crucial information on the integrated spectral properties of the remnant age distribution, which in turn can be used to test models of remnant lobe evolution, as discussed in Section 4.3.2.

We perform the selection using the LOFAR images and catalogues at both high and low resolution, combined with higher frequency surveys (NVSS, WENSS, FIRST). In particular, for each selection criterion we use different combinations of catalogues and images and different flux limits to obtain the best selection. This means that our selected samples should be considered as independent (partially overlapping) datasets and the overall completeness of our remnant candidates’ search cannot be established.  A description of the different selection methods is presented in the following sections.

\subsection{Ultra-steep spectral index selection}
\label{uss}

Spectral ageing models predict the integrated radio spectrum of active sources to be a broken power law. Values in the range 0.5-0.7 are classically assumed for the spectral injection index $\rm \alpha_{inj}$ (e.g. \citealp{blandford1978}) below a break frequency $\rm \nu_{break}$, while a spectral index equal to $\rm \alpha= \alpha_{inj}+0.5$ is expected above $\rm \nu_{break}$ according to the continuous injection model (\citealp{jaffe1973}; \citealp{carilli1991}). After the active nucleus of the radio galaxies switches off, the spectrum steepens well beyond this value. For this reason we consider  all sources having $\rm \alpha_{150}^{1400} > 1.2$ to be good candidates (\citealp{komissarov1994}). This value allows us to collect the largest number of remnant candidates while minimizing contamination from active steep sources. False positives are particularly expected to come from active Fanaroff-Riley class II (FRII, \citealp{fanaroff1974}) radio galaxies for which recent observations indicate high values of injection spectral indices $\rm \alpha_{inj}\gtrsim 0.7$ \citep{harwood2016}. However, FRII radio galaxies are not expected to dominate our sample, as shown in Section \ref{skad}. 

In the L45N catalogue we find that 22 sources (4.1\%) have $\rm \alpha_{150}^{1400} > 1.2$. These include the nine sources without a NVSS detection. When accounting for the errors on the spectral indices, this percentage can vary in the range [3.7\%-6.3\%].

\subsection{Spectral curvature selection}
\label{spc}

The spectral curvature, defined as $\rm SPC = \alpha_{high}-\alpha_{low}$ (in this case $\rm \alpha_{low}=\alpha_{150}^{325}$ and $\rm \alpha_{high}= \alpha_{325}^{1400}$), is introduced to select sources whose global spectrum is not steep enough to be included in the ultra-steep sample but show a significant curvature due to particle ageing. 

\begin{figure}
\begin{center}
{\includegraphics[angle=0, width=10cm]{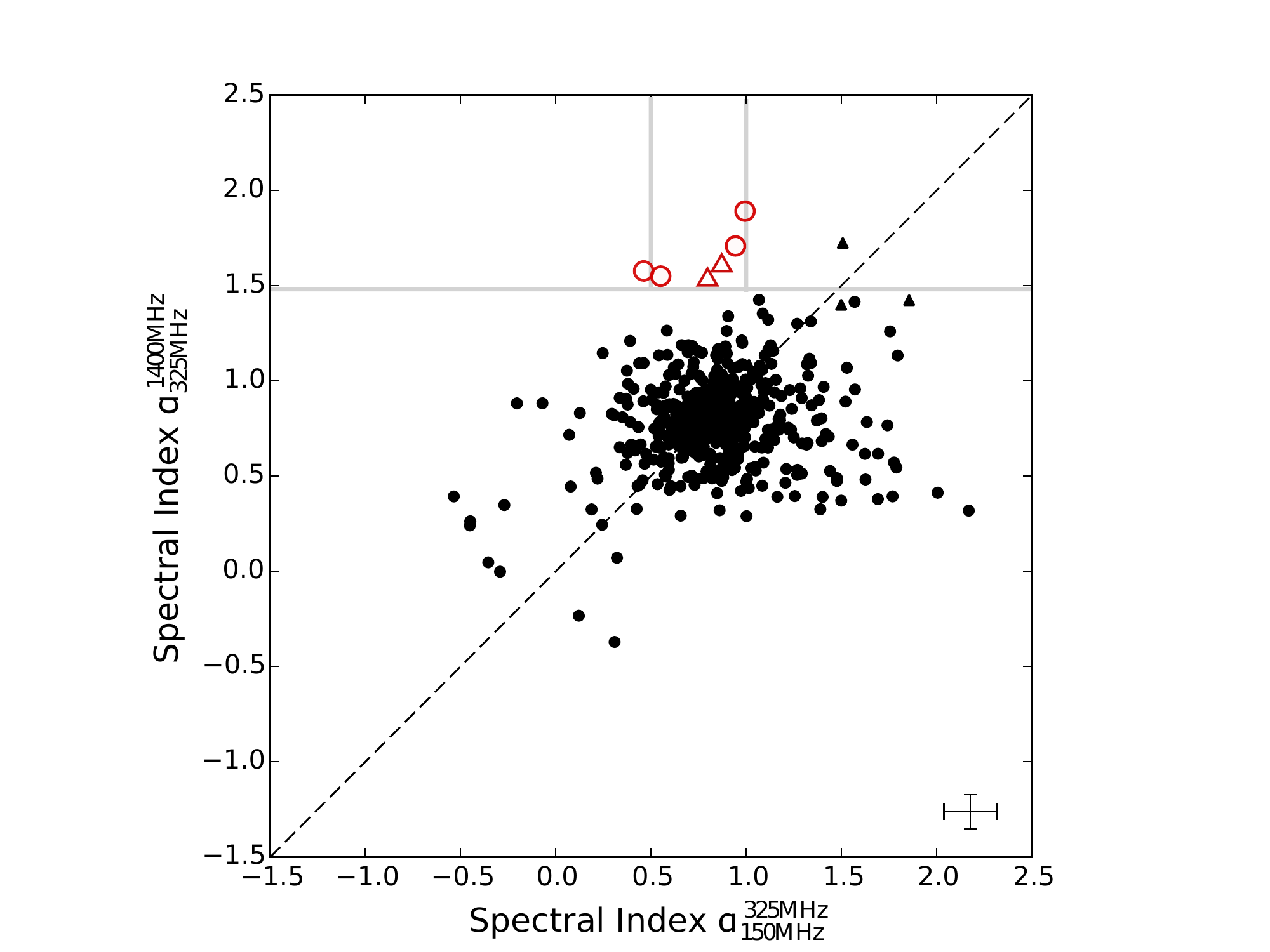}}
\caption{Radio colour-colour plot for sources in the L45NW sample. Triangles represent sources that are not detected in NVSS. Sources that have been selected using the spectral curvature criterion described in Section \ref{spc} are marked with open red symbols. In the bottom right corner a mean error for the points in the plot is show. A black dashed line represents the 1:1 diagonal. Grey lines represent the constraints used for the selection.}
\label{fig:spc1}
 \end{center}
\end{figure}

In Fig. \ref{fig:spc1} we show the spectral indices between 150 and 325 MHz compared to the spectral indices between 325 and 1400 MHz derived in the L45NW. Triangles indicate sources that do not have a detection at 1400 MHz and therefore only have a lower limit on the spectral index $\rm \alpha_{325}^{1400}$. We can see that most of the sources cluster around the 1:1 diagonal indicating a straight  power-law spectrum, within the errors.

For our search we consider as good candidates all those sources that have, within the errors, $\rm 0.5\leq\alpha_{150}^{325}<1$, the typical range for active sources, and $\rm \alpha_{325}^{1400} \geq1.5$. The constraint in $\rm \alpha_{150}^{325}$ is chosen to avoid sources that are either already ultra-steep in the range 150-325 MHz or that have a turnover at low frequencies due to self-absorption processes, while that in $\rm \alpha_{325}^{1400}$ is chosen to select ultra-steep sources at the highest frequencies available. By applying these selection criteria we identify six sources (out of 453) which are shown in Figure \ref{fig:spc1} as red open symbols.

\bigskip

\subsection{Morphology selection}
\label{morpho}

Morphological selection can be used as another useful tool to identify remnant sources (e.g. \citealp{saripalli2012}). It potentially allows us to recognize remnants whose spectrum is still not ultra-steep or curved and thus are missed by the spectral selection. 

Unequivocally defining the morphology of a remnant radio galaxy is challenging. Indeed, the shape of the source at the end of its life depends on its original morphology and physical properties and  on the conditions of the surrounding medium. The classical prototype of a remnant radio galaxy is thought to have relaxed morphology without compact components like a core, hot-spots, or jets. Moreover the shape of the remnant plasma may become amorphous due to expansion if the source is over-pressured at the end of its life (\citealp{blundell1999}; \citealp{wang2008}). 

In light of this, we perform the morphology selection using the LOFAR high-resolution image via visual inspection. The low-resolution image, which is more sensitive to large-scale emission, is also used to confirm the candidates. For the selection we use the following criteria: (i) extended size in the high-resolution map ($\rm \gtrsim 60 \ arcsec$ equal to $\rm \sim3\times beam$ to to enable a visual inspection of the shape), (ii) relaxed morphology with low surface brightness (< 50 mJy $\rm arcmin^{-2}$) at 150 MHz, (iii) absence of compact features (core, jets, or hot-spots) in the LOFAR image, and (iv) absence of compact features above 3$\sigma$ in the FIRST at 5 arcsec resolution. Our final sample of morphologically selected candidate remnant radio galaxies is composed of 13 sources. According to the source size measurements made by PyBDSM in the high-resolution catalogue, there are 69 sources with a major axis >60 arcsec. This is considered a lower limit since some of the multi-component extended sources may not be automatically combined  by the source extraction software. The candidate remnant radio source fraction among sources with sizes exceeding 60 arcsec is therefore <13/69 ($\rm \lesssim 20\%$). The LOFAR contours of the 13 candidates overlaid on the FIRST maps are shown in Fig. \ref{fig:morpho} marked  `M'.

\subsection{Low radio core prominence selection}
\label{cp}

With the morphological criteria presented in Section \ref{morpho} we have selected sources with relaxed shapes and lacking any compact components. However, there may be remnant sources without core radio emission but where the hot-spots are still visible if the jets have recently (less
than a jet travel time) switched off  (e.g. 3C28 \citealp{feretti1984}, \citealp{harwood2015}).

To identify these candidate remnant radio sources, we perform a further selection using the core prominence, i.e. the ratio between the core power at high frequency and the power of the extended emission at low frequency. \cite{deruiter1990} show that the core prominence in the B2 sample (\citealp{colla1970}, \citealp{fanti1978}) is inversely proportional to the radio luminosity of the source varying in the range 0.1-0.001 for radio powers in the range $\rm 10^{24}-10^{26} \ W Hz^{-1}$. In agreement with this finding, the objects in the 3CRR sample, which contains the most powerful radio galaxies in the sky, show a mean radio core prominence of $\sim$3$\rm \times 10^{-4}$ (\citealp{giovannini1988}, \citealp{mullin2008}). We  therefore expect remnant radio galaxies to have, on average, a CP$\rm \lesssim \times 10^{-4}$.

For this selection we follow the approach taken by \cite{hardcastle2016}. We use the FIRST to search for any visible core emission. We initially consider all sources with the following characteristics: (i) total flux density at 150 MHz above 90 mJy (at 45 arcsec) to be able to put the tightest possible upper limit to the CP in case of core not detection (<0.005) and (ii) size above 40 arcsec (in the LOFAR map at 18 arcsec resolution) to allow the visual identification of the radio core.

Using these criteria we are left with 34 sources for which we visually checked the presence of a core in the FIRST images. We find that 10 out of 34 sources clearly do not show core emission down to 3$\sigma$ and that can be considered candidate radio remnants ($\rm \sim$30\%). As expected, there is some overlap with the morphology selection presented in Sec. \ref{morpho}. In particular three sources are identified with both CP and morphology criteria (J103414+600333,  J104732+555007, and J105230+563602). For all the candidates we compute an upper limit of the CP as the ratio between the 3$\sigma$ level in the FIRST image (at the local noise) and the total LOFAR flux density, giving an average value of $3.3\times10^{-4}$. One caveat of this method is that it may be affected by  core flux density variability if the flux density of the core and the extended structure do not come from simultaneous observations. The values of the computed core prominence are listed in Table \ref{tab:morpho}. The LOFAR contours of the ten candidates overlaid on the FIRST maps are shown in Fig. \ref{fig:morpho} marked  `CP'.

The candidate percentage should be considered as an upper limit for two reasons. Firstly, the low sensitivity of the FIRST survey does not allow us to put tight constraints on faint radio cores and future high-sensitivity and high-resolution observations may reveal the presence of a faint radio core for some sources. Secondly, the total fraction of extended radio galaxies may be underestimated because some multi-component sources may be missed by the source extractor software as already mentioned in Sec. \ref{morpho}.

\subsection{Results of remnant radio galaxy selection process}
\label{resultsobs}

\begin{small}
\begin{table*}[h]
        \centering
        \caption{Final list of remnant radio galaxy candidates selected with different criteria (see Section \ref{selection}). In Col. 1 are the source names in J2000 coordinates;  Col. 2 the source component (N=North, S=South, E=East, W=West);  Col. 3 the flux densities at 150 MHz from the L45N catalogue;  Col. 4 the spectral indices between LOFAR 150 MHz and NVSS 1400 MHz;  Col. 5 the radio core prominence (CP) computed as the ratio between the FIRST 1400 MHz core flux density and the LOFAR 150 MHz total flux density; and  Col. 6 the selection methods used to identify the source (M=morphology, CP=low core prominence, US=ultra-steep spectrum with size>40 arcsec). Sources showing ultra-steep spectra are marked with an asterisk.}
        \small
                \begin{tabular}{*6l}
                \hline
                \hline
            Name   & Component & $\rm S_{150MHz} $&$\rm \alpha_{150}^{1400}$ & CP & Selection \\
            &&[mJy]&&&criteria\\
            \hline
                \hline
            J102818+560811 & total&50.2&0.68 & <$\rm 2.9\times10^{-3}$& M\\
            J102842+575122 & total & 32.3&-&<$\rm1.3\times10^{-2}$& M\\ 
                & N & 15.0 &>0.88 & \\
                 &S & 17.3 &>0.94 &\\
                J102905+585721 & total&34.3&>1.2* & <$\rm3.5\times10^{-3}$& M\\
                J102917+584208&total&112.3&     0.95&   <$\rm4.3\times10^{-4}$& CP\\
                J103132+591549&total&121.3&     1.16    & <$\rm3.8\times10^{-4}$& CP\\
                J103414+600333 & total &151.7& - &<$\rm8.9\times10^{-4}$& M, CP \\
                 &NE &24.4& 0.47 & \\
                 &SW &127.3& 1.00 & \\
                J103602+554007 &total& 27.5&>1.01 & <$\rm2.1\times10^{-3}$& M\\
                J103641+593702 &total& 30.5&>0.99 & <$\rm1.9\times10^{-3}$& M\\
                J103805+601150&total&181.6&     0.90&   <$\rm2.6\times10^{-4}$&CP\\
                J104208+592030&total&152.5&     1.09&   <$\rm3.7\times10^{-4}$&CP\\
                J104516+563148 & total &25.8&-& <$\rm3.4\times10^{-3}$& M\\
                &E &15.9& 1.11 & \\
                 &W &9.9& 0.94 & \\
                J104646+564744&total&758.7&     0.94&   <$\rm6.6\times10^{-5}$&CP\\
                J104732+555007 & total & 91.2&-&<$\rm9.8\times10^{-4}$& M, CP\\
                    & NW &42.6& 1.25* & \\
                        & SE&48.6 & 1.20* &\\
                J105230+563602 &total &204.2&0.87&<$\rm7.4\times10^{-4}$& M, CP\\
                J105402+550554 &total&46.4 &>0.86& <$\rm9.6\times10^{-4}$& M\\
                J105554+563532 & total &83.9 & 1.43* & <$\rm1.4\times10^{-3}$ & US\\
                J105703+584721&total&1518.3&    0.91&   <$\rm4.4\times10^{-5}$&CP\\
                J105729+591128 &total&  11.9 &1.08 & <$\rm3.7\times10^{-3}$& M\\
                J110108+560330 & total & 58.2 &1.22* & <$\rm2.3\times10^{-3}$ &US\\
                J110136+592602 & total &45.7&- & <$\rm2.6\times10^{-3}$&M\\
                &NE &22.0& 1.14 & \\
                &SW &23.7 &1.17 & \\
                J110255+585740 & total & 97.5 &1.40* & <$\rm1.2\times10^{-3}$ & US \\
                J110420+585409&total&225.9&     0.84&   <$\rm1.7\times10^{-4}$&CP\\
                J110806+583144&total& 27.1&0.80&  <$\rm1.6\times10^{-3}$& M\\
                \hline
                \hline  
                \end{tabular}
                \label{tab:morpho}
\end{table*}
\end{small}

\begin{figure*}[b]
\centering
{\includegraphics[width=16cm]{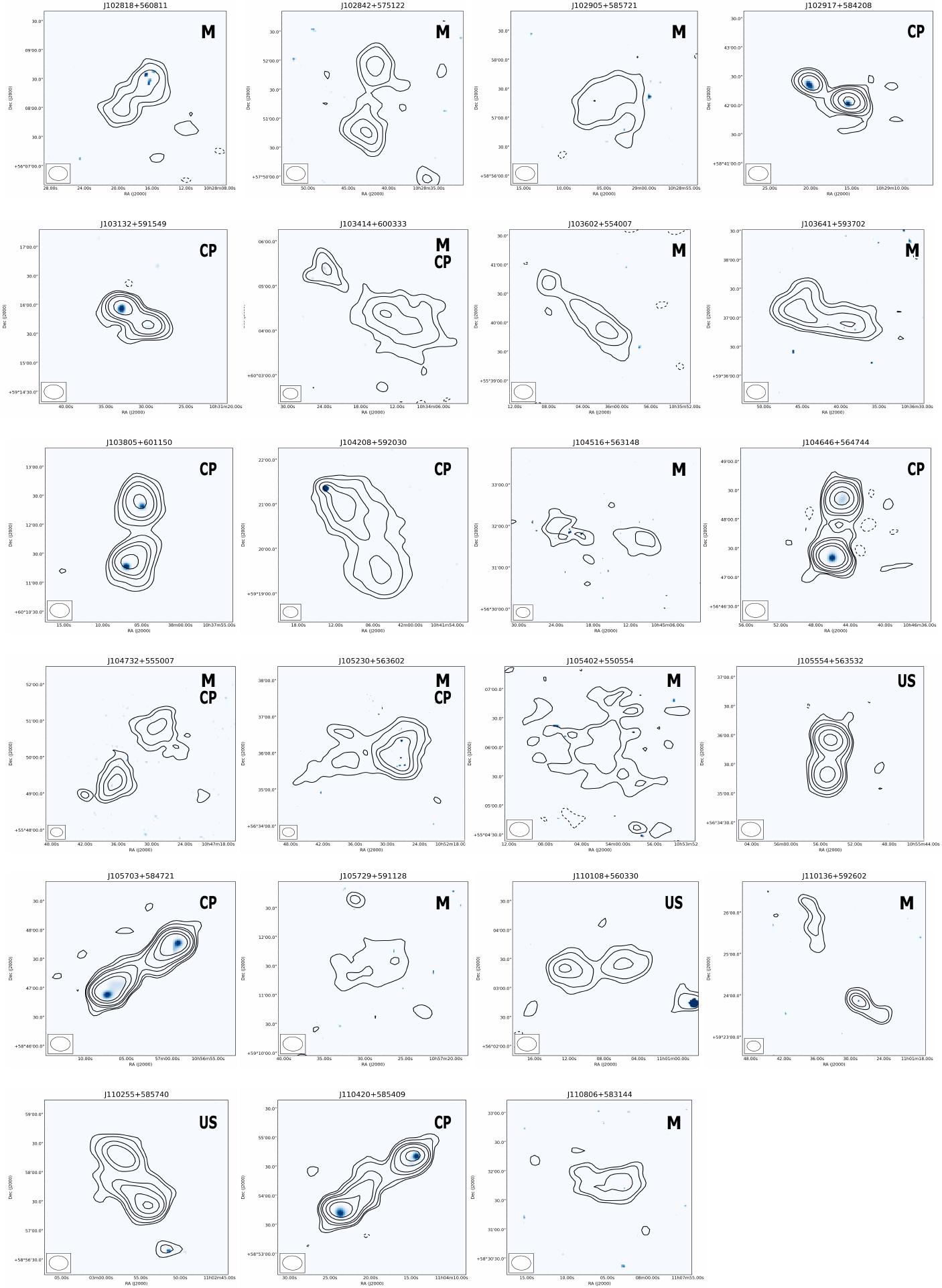}}
\caption{Candidate remnant radio galaxies selected on the basis of different selection criteria (see Sect. \ref{selection}). LOFAR radio contours (-3, 3, 5, 10, 15, 30, 50 $\rm \times \ \sigma_{local}\approx0.5\ mJy$) of the highest resolution map $18\times15$ arcsec are overlaid on the FIRST map, whose scale is set in the range [3$\rm \sigma_{local}-S_{peak}$]. The LOFAR beam is shown in the bottom left corner. The selection criterion used  to identify the source is shown in the top right corner (M=morphology, CP=low core prominence, US=ultra-steep spectrum with size>40 arcsec). }
\label{fig:morpho}
 \end{figure*}

In this section we investigate the results from the selections presented above.

With the ultra-steep spectrum criterion (see Section \ref{uss}), which is a widely used technique in the literature to search for remnants, we have selected 4.1\% [3.7\%-6.3\%] of the sources in the catalogue. Although the L45N catalogue includes extended sources (>40 arcsec) that were excluded in the Lockman-wide catalogue by M16, the resulting percentage is comparable to the value they found (4.9\%). This suggests that the ultra-steep spectral indices are not preferentially associated with very extended sources. The same spectral study performed on the Lockman-WSRT at much higher sensitivity and resolution results in a consistent percentage of 6.6\%, varying in the range [4.3\%-9.2\%]. We mention here that the fraction found in the Lockman-WSRT catalogue by M16 is consistent with previous analysis performed by \cite{afonso2011} and \cite{ibar2009} at comparable flux limits on a smaller portion of the same field ($\rm 0.56 \ deg^2$) at higher frequencies ($\rm 6.3\%, \  \alpha_{625}^{1400} >1.3$). Studies on different fields still find low fractions of ultra-steep spectrum sources, though a detailed comparison is difficult due to the varying observational characteristics (frequencies, sensitivities, field of view). 
For example, \cite{cohen2004} using the VLSS catalogue with a flux limit of $\rm S_{74MHz}$=0.7 Jy $\rm beam^{-1}$ find 2.7\% of sources with $\rm \alpha_{74}^{325} >1.2$, and \cite{sirothia2009} using dedicated observations on a field of $\rm 0.25 \ deg^2$ and a flux limit of $\rm S_{153MHz}$=2.5 mJy $\rm beam^{-1}$ find 3.7\% of sources with $\rm \alpha^{1260}_{250}>1.03$). 

It is important to stress that our sample of ultra-steep spectrum sources not only includes remnant radio galaxies but also other classes of sources with the same spectral properties. Among these are cluster relics and halos (e.g. \citep{vanweeren2009}) and mostly high-z radio galaxies (HzRGs, e.g. \citealp{roettgering1994}, \citealp{debreuck2000}). In our sample of ultra-steep spectrum sources we have one source which is part of the cluster A1132 and which will be excluded from the further analysis due to its uncertain nature. 

Disentangling remnants from HrRGs is difficult without the optical identification of the host galaxy, and we defer this analysis to a future study. A tentative discrimination between the two classes of sources can be done using the source angular size. We restrict our sample of ultra-steep spectrum remnants to sources having a deconvolved angular size >40 arcsec in the LOFAR high-resolution catalogue. At redshift $z=2$ this would correspond to a linear size >300 kpc, which is unlikely for a HzRG (\citealp{ker2012}) and supports the classification as remnant radio galaxy at lower redshift. By using this further criterion we select 3/21 ultra-steep sources as good candidate remnants, which  represent only <1\% (3/543) of all sources in the L45N catalogue. This should be considered a conservative upper limit since some of the compact sources may be small remnants. The LOFAR contours of the three candidates overlaid on the FIRST maps are shown in Fig. \ref{fig:morpho} marked  `US'.

The selection based on the SPC is used to include radio galaxies in an intermediate evolutionary stage (see Section \ref{spc}) where the spectrum starts to be curved because of the particle ageing, but is still not ultra-steep down to low frequencies. With this method we identify six sources. However, in order to avoid contamination from HzRG, we apply here the same angular size criterion described in the previous paragraph. Because all six sources are unresolved at 18-arcsec resolution we reject them all from our selection. Moreover, we note that all six sources have already been identified by the ultra-steep spectrum criterion $\rm \alpha_{150}^{1400} >1.2$. Because the steepening of the spectrum at early stages occurs mostly at higher frequency, we suggest
 that  including observations at 5000-MHz or higher in this kind of analysis is essential for the selection to produce complementary results to the ultra-steep spectrum method (see e.g. \citealp{brienza2016}). In our work, sources that have spectral breaks at frequencies >1400 MHz are missed by the SPC selection and sources that have very curved spectra also have ultra-steep spectral indices at low frequencies and are therefore already included  in the ultra-steep selection criterion. Complementary data at such low flux limits at 5000 MHz are not currently available over large fractions of the sky. Observations of the Lockman Hole at 15000 MHz have been carried out (\citealp{davies2011}; \citealp{whittam2013}), but due to their very low sensitivity they do not provide any useful constraints, so we have not included them in this work.

Using the morphological criteria described in Section \ref{morpho} we have selected 13 remnant candidates. It is worth mentioning that by following these criteria there can still  be contaminations from other class of sources, for example face-on spiral galaxies or cluster halos, which need to be identified and manually removed from the sample. In this specific search we  identified the spiral galaxy NGC~3445 and excluded it from further analysis. Of course we cannot exclude the possibility that some candidates may just be active with atypical morphologies, and future observations are planned to verify this possibility. Typical values of surface brightness at 150 MHz in these  morphologically selected sources are in the range 10-30 mJy $\rm arcmin^{-2}$, comparable to what has been found for the source Blob1 by \cite{brienza2016} (15 mJy $\rm arcmin^{-2}$). For these sources we also compute an upper limit to the core prominence as the ratio between the FIRST 1400 MHz core flux density $\rm 3\sigma$ upper limit and the LOFAR 150 MHz  total flux density in the catalogue at 45 arcsec (see Table \ref{tab:morpho}). We find upper limits in the range $\rm 7.9\times10^{-4}-1.3\times10^{-2}$ which do not provide tight constraints. We also derive the spectral index $\rm \alpha_{150}^{1400}$ using the LOFAR flux density at 45 arcsec resolution and the NVSS (see Table \ref{tab:morpho}). When the NVSS map does not show a source detection, we compute a 3$\rm \sigma$ upper limit by measuring the standard deviation of the flux density in ten different boxes surrounding the source location. The results of this computation show that only two sources have spectral indices $\rm \alpha_{150}^{1400}>1.2$ (J102905+585721, J104732+555007) and four sources only have lower limits (J102842+575122, J103602+554007, J103641+593702, J105402+550554). Interestingly, only a fraction of  these sources, <46\% (6/13 sources, three of which  have upper limits in spectral index),  are found to have $\rm \alpha_{150}^{1400}$>1.2.

With the low radio core prominence criterion (see Section \ref{cp}) we  selected a sample of 10/34 sources whose central radio cores appear to be inactive while the hot-spots may still present. Three of these sources have been already selected with the morphology criteria described above. The observed fraction is consistent with that found in the H-ATLAS field by \cite{hardcastle2016}. Also for this sample, we computed the spectral index $\rm \alpha_{150}^{1400}$ using the LOFAR map at 45 arcsec resolution and the NVSS map and we find that only a low fraction ($\sim$7\%) have ultra-steep  spectra with $\rm \alpha_{150}^{1400}$>1.2.

To summarize, we have selected 23 remnant radio galaxy candidates: 3 with the USS criterion and a conservative size cut at 40 arcsec, 13 with the morphology criterion, and 10 with the low core prominence criterion (3 sources are selected with both the morphology and the low core-prominence criterion). A list of the sources with their respective properties can be found in Table \ref{tab:morpho}. Radio maps of these sources are shown in Figure \ref{fig:morpho}.

\subsection{Implications from the selection}

The four empirical methods described in the previous sections have allowed us to select 23 candidate remnants with very different characteristics. We suggest that the different classes of objects found are related to different stages of the radio galaxy evolution.

The ultra-steep spectrum selection is expected to be strongly biased towards very aged plasma, i.e. very evolved remnant sources, which should only represent a subclass of the entire remnant population as shown in \cite{godfrey2017} and further investigated in the second part of this paper. Only three ultra-steep spectrum remnant candidates have been identified from the L45N sample. 

Identifications based on morphology are crucial to expanding the selection to include younger remnants. Sources selected via this method are expected to follow an approximately uniform distribution in remnant age. Therefore, the fraction of morphologically selected remnant candidates that have ultra-steep spectra is an indication of the age distribution in the remnant population. Interestingly, only a fraction of these sources, <46\% (six sources, of which three have upper limits in spectral index),  are found to have $\rm \alpha_{150}^{1400}$>1.2. The observed trend suggests that ultra-steep spectrum remnants in the range 150-1400 MHz represent only a fraction of the entire remnant population, and that remnant plasma undergoes a very rapid luminosity evolution. This is   investigated further in the second part of the paper. 

Finally, candidate remnants selected on the basis of low core prominence are thought to be the youngest remnant radio galaxies that we can select where the time elapsed since the core switch off is less than the jet travel time to the edges of the sources. This is consistent with the fact that only $\sim$10\% of these sources are found to have ultra-steep spectra with $\rm \alpha_{150}^{1400}$>1.2 suggesting the plasma has not yet aged significantly. 
 
Deep high-resolution imaging at complementary radio frequencies are planned to further investigate the properties and confirm the nature of the 23 candidates. Optical identification of the host galaxies together with redshift information will give us insights into the remnants' surrounding environment and allow us to apply spectral ageing models to derive the age of the plasma.

The main limitation of the selected candidates is that they do not represent a complete sample; therefore, we cannot directly compare the results  of each selection criterion. This restriction will be overcome soon as new surveys at high sensitivity and higher frequency are released, allowing for a full exploitation of the LOFAR data. APERTIF (\citealp{oosterloo2009}) is going to provide deep ($\rm \sigma$=0.1 mJy) 1400 MHz data at comparable resolution to LOFAR, allowing a systematic search of ultra-steep spectrum remnants down to low flux limits. The VLA Sky Survey (VLASS, \citealp{myers2014}) will instead provide maps of the northern sky at 3000 MHz with 0.12 mJy flux limit and 2.5 arcsec resolution. This will expand the low core prominence selection down to low limits.

\section{Simulating the population of active and remnant FRI radio galaxies}

For a long time there have been claims that sensitive low-frequency surveys will lead to the discovery of many remnant radio galaxies (e.g. \citealp{rottgering2006}, \citealp{kapinska2015}) and LOFAR now gives us the opportunity to investigate whether this is the case.
However, precise modelling of the evolution of remnants  and predicting the number of remnants in the radio sky remains challenging. Recently, \cite{godfrey2017} presented a study based on a VLSS-selected sample and mock catalogues of high-power radio galaxies to derive the fraction of remnants expected in flux limited samples. 

Following this, and in parallel to our empirical search, we have also simulated catalogues of low-power radio galaxies (see Section \ref{skad}) to provide constraints on the number of remnants expected in the Lockman Hole field and more generally in the LoTSS. As shown below, due to the higher sensitivity of the LOFAR data with respect to the VLSS used by \cite{godfrey2017}, both low-resolution and high-resolution radio catalogues are dominated by low-power sources instead of high-power ones. Therefore, an extension of \cite{godfrey2017} simulations to low-power radio galaxies is required if we want to consistently compare  the results of the simulations with those from these Lockman Hole observations.

To create mock catalogues, we follow the same approach proposed by \cite{godfrey2017} who use the Monte Carlo method to simulate a flux limited sample of FRII radio galaxies. This allows us to use available analytical spectral evolution models and observed source properties to derive, to the first order, the fraction of remnants in flux limited samples, given a set of model assumptions. With this approach we can directly compare the empirical catalogue with the mock one by applying the same flux density cut, and compare the results by applying the same selection criteria. The main limitation is that the mock catalogues that are generated only contain information on the spectra of the sources and not on the  morphology, so a direct comparison can only be made based on the spectral criteria. In particular we focus on the ultra-steep spectral index criterion.
Given that the fraction of ultra-steep spectral index sources found in the Lockman-WSRT and in the L45N catalogues are similar,  as a reference
for the following simulations we  use  the Lockman-WSRT catalogue, which is the deepest catalogue currently available with a mean flux limit of 1.5 mJy.

\subsection{Dominant classes of radio source in our sample: SKADS Simulations}
\label{skad}

In order to investigate quantitatively  the dominant population of radio sources in the catalogues described in Sect. \ref{data}, we use the SKADS Simulated Skies (S3, \citealp{wilman2008}). We run two different simulations to reproduce both the Lockman-WSRT catalogue and the L45N catalogue as follows:

\begin{itemize}
\item{\textit{L45N} - 30 sq. degrees with flux density detection limit at 151 MHz $\rm S_{151,min}=40$ mJy. This predicts 602 sources (vs 543 in the observed field) of which 69\% are FRI, 28\% are FRII, 2\% are Gigahertz-Peaked Spectrum, 1\% are radio quiet AGN, and 1\% are star forming galaxies.}

\item{\textit{Lockman-WSRT} - 6.6 sq. degrees with flux density detection limit at 151 MHz $\rm S_{151,min}=1.5$ mJy. We choose 1.5 mJy as an average 5$\sigma$ value throughout the field. This predicts 1388 sources (vs 1376 in the observed field) of which 67\% are FRI, 2\% are FRII, 3\% are Gigahertz-Peaked Spectrum, 13\% are radio quiet AGN, and 15\% are star forming galaxies }
\end{itemize}

In both samples we can see that the predominant sources ($\rm \sim$70\%) are FRIs, i.e. typically low-power radio galaxies. The main difference between the two samples is the number of FRII radio galaxies, which drastically increases at higher flux densities, and in the number of star forming galaxies, which increases at low flux limits. Therefore, we restrict the following simulations to FRI radio galaxies which are expected to represent the bulk of the population in our empirical samples. In particular, we only model FRIs with lobed morphology (also referred to as `bridged'), which constitute $\rm \sim$62\% of the B2 catalogue of low-power radio galaxies according to \cite{parma1996}. This choice is justified by the absence of models in the literature describing the dynamical evolution of `naked-jet' or `tailed' FRI due to the complexity of the physics involved. We  therefore expect $\rm \sim 43$\% of our empirical catalogue to be constituted of lobed FRIs, which is what we   model in this section. In particular, we know that 6.6\% of the sources in the Lockman-WSRT catalogue are found to be ultra-steep spectrum sources. We can therefore put an upper limit on the fraction of remnant lobed FRIs with ultra-steep spectrum equal to < 0.066/0.43 $\sim$15\%, which can vary in the range [10\%-21\%] according to errors.

\subsection{Simulation approach}
\label{simulations}

To create mock catalogues of radio galaxies we simulate several thousands of sources using a Monte Carlo approach based on radio galaxy evolution models. In particular, we present two simulations based on two different evolution models described later in this section:  radiative evolution only (CI-off, \citealp{komissarov1994}) and  radiative and dynamical evolution (\citealp{komissarov1994} and \citealp{luo2010}). Both models depend on a set of parameters that describe the source properties and the surrounding environment (see Table \ref{tab:param}). Some of these parameters are kept fixed, while some others are sampled from probability distributions based on empirical observations of low-power radio galaxies. All parameters are treated as independent variables in the simulation. The details of the assumptions made for each parameter are  discussed below. We calculate the radio spectra using the code presented in \cite{godfrey2017}, which is based on the synchrotron model proposed by \cite{tribble1991, tribble1993}, and further expanded to an implementable form by \cite{hardcastle2013} and \cite{harwood2013}. According to this, the magnetic field within each volume element of the lobe is a Gaussian random field, with varying magnetic field orientation and magnitude. We refer the interested reader to \cite{godfrey2017} for a full description of the  implementation of the model. Here we describe the main steps used to  generate the mock catalogues:

\begin{itemize}
\item set the number of sources to be generated (several thousands);
\item set the values of the fixed parameters of the model and sample the other parameters from the corresponding probability distributions (see Figure \ref{tab:param});
\item for each source, given its set of parameters, compute an upper limit to the flux density at the sample selection frequency (150
MHz) following Section 4.6 in \cite{godfrey2017};
\item apply a flux density cut consistent with the deepest available observations (1.5 mJy) so that all sources below the threshold are rejected;
\item for the remaining sources, calculate the model radio galaxy spectrum accurately using numerical integration of equations 1 and 9 in \cite{godfrey2017};
\item derive flux densities at the observed frequencies and compute relevant spectral indices;
\item reject all sources for which the accurate flux density at the selection frequency is below the flux limit.
\end{itemize}
In what follows, we describe each of the model parameters and its corresponding probability distribution or fixed value.

\subsubsection*{\textbf{Redshift} z }
\label{redshift}

To sample the redshifts we use a probability distribution of the form

\begin{equation}
p(z) \propto  \rho_v (z)  \frac{dV}{dz}
,\end{equation}

\noindent where $\rm \rho_v(z)$  is the volume density of radio galaxies as a function of redshift and $\rm \frac{dV}{dz}$ is the differential comoving volume element of a spherical shell. The comoving volume element for a flat Universe ($\rm \Omega_k=0$) is derived following \cite{hogg1999}. Following the luminosity functions found by \cite{willott2001} and \cite{wilman2008} for low-power sources (which are also used in the SKADS simulations), we consider $\rm \rho_v(z)$ to be a piece-wise power law:

\begin{equation}
\rm \rho_v(z) \propto \begin{cases}(1+z)^k, \ \ \ \ k=4.3 & if \ \ \ \ z<z_{l0}\\
(1+z_{l0})^k, \ \ k=0 & if \ \ \ \ z\leq z_{l0}\leq  5\\
\end{cases}
\end{equation}

\noindent with $\rm z_{l0}=0.706$. 

We note here that the effect of cosmological luminosity evolution is not considered in this work. This is justified by the fact that FRI sources do not show a strong cosmological evolution (e.g. \citealp{wall1997}).

\subsubsection*{\textbf{Jet power \boldmath$Q_{jet}$\unboldmath}}
\label{qjet}

For the jet power probability distribution we assume a power-law distribution with slope $\rm \delta=0.6$ (\citealp{willott2001} and \citealp{kaiser2007}) in the range $\rm 10^{34}-5\times 10^{37} W $ (\citealp{luo2010}). 
  
\subsubsection*{\textbf{Active time \boldmath$t_{on}$}\unboldmath}
\label{ton}

The length of the active time of radio galaxies is still not very well constrained, especially for FRI radio galaxies. Even though we are aware of the discrepancies between spectral and dynamical ages found in literature (e.g. \citealp{eilek1996}),  we use here spectral ages as a reference to be consistent with our spectral modelling. For the B2 sample, radiative ages were calculated by \cite{parma1999} using the model by \cite{jaffe1973}. The mean source age of the sample is 31 Myr, while the maximum and minimum values are 5 Myr and 75 Myr, respectively. In our simulation we sample $\rm t_{on}$ from a truncated normal distribution with $\rm t_{on,min}=1 \ Myr, t_{on,max}=300 \ Myr $, mean $\rm \overline{t_{on}} = 40\ Myr, \ and \ standard \ deviation \ \sigma_{t_{on}}=30\ Myr$. The parameters of the distribution have been selected so that the final age distribution in the mock catalogue reproduces the age distribution of the B2 sample. 

\subsubsection*{\textbf{Source age \boldmath $t_{obs}$}\unboldmath}
\label{tobs}

We sample the source ages from a uniform distribution with $\rm t_{obs, min} = 0.1$ Myr and $\rm t_{obs, max}=400 \ Myr$. 

\subsubsection*{\textbf{Particle energy injection index p}}
\label{alpha}

The particle injection index of each radio source is sampled from a truncated normal distribution with $\rm p_{min}=2.0, \ p_{max}=2.4 $, mean $\rm \overline{p} = 2.2, \ and \ standard \ deviation \ \sigma_p=0.2$. This corresponds to a continuous distribution of injection spectral index in the range 0.5 < $\rm \alpha$ < 0.7 peaking at 0.6, which is  in agreement with the empirical results by \cite{laing2013}.

\subsubsection*{\textbf{Electron energy fraction \boldmath$\epsilon_e$}\unboldmath}
\label{electron}

The electron energy factor $\rm \epsilon_e$ represents the fraction of jet power that is converted to the internal energy of the relativistic electrons. In the model we assume that the internal energy of the radio source is equally partitioned between particles and magnetic energy. We assume that the electron-to-proton ratio is  k=1, as suggested by \cite{croston2008} for lobed FRIs. This translates into an electron energy factor of $\rm \epsilon_e=0.25$.

\begin{small}
\begin{table*}[h!]
        \centering
        \caption{List of the parameters used in the simulations described in Section \ref{simulationsfull} with their respective references. References: (1) \cite{luo2010}; (2) \cite{laing2013}; (3) \cite{croston2008}; (4) \cite{willott2001}; (5) \cite{parma1999}; (6) \cite{wilman2008}. *Magnetic fields are computed assuming pressure balance between the source lobes and the external environment for each source, as described in Section 4.2.2}. 
        \small
                \begin{tabular}{*4c}
                \hline
                \hline
                &&&\\
                Parameter& Radiative model only & Radiative and dynamical model & Reference\\
        &&&\\           
                \hline
                \hline
                &&&\\
                $\rm \beta$ - Power index of the ambient medium density profile & - &1.5 & 1\\
                $\rm r_0$ - Normalization radius of the ambient medium density profile [kpc] & -& 2 & 1\\
                $\rm p_0$ - Normalization pressure of the ambient medium density profile [Pa] & -& $\rm 3\times 10^{11}$ & 1\\
           $\rm c_s$ - Sound speed [km/s] & -&1000 & -\\                
                $\rm \chi$ - Geometric factor & - & 1& 1\\
                $\rm \Gamma$ - Adiabatic index & - & 4/3 & 1 \\  
                $ \rm \gamma_l$ - Minimum Lorentz factor & 5& 5 & 1\\
                $ \rm \gamma_m$ - Maximum Lorentz factor & $\rm 1\times10^{7}$ & $\rm 1\times10^{7}$ & 1\\
                p - Relativistic particle spectral index &2.0-2.4 &2.0-2.4& 2\\
                $\rm \epsilon_e$ - Electron energy fraction &0.25 &0.25& 3\\
                $\rm Q_{jet}$ - Jet power [W] &  $\rm10^{34}-5\times10^{37}$ &  $\rm 10^{34}-5\times10^{37}$& 1, 4\\
                B - Magnetic field [$\rm \mu$G] & 5 & 0.02-50* & 5\\
                $\rm t_{on}$ - Duration of the active phase [Myr] &1-300 &1-300 & 5\\
                $\rm t_{obs}$ - Source age [Myr] & 0.1-400 & 0.1-400 & -\\      
                z - Redshift&0-5 &0-5 &4, 6\\   
                &&&\\   
                \hline
                \hline  
                \end{tabular}
                \label{tab:param}
\end{table*}
\end{small}

\textbf{\subsubsection{Simulation with radiative evolution alone}}
\label{simulationsrad}

In this simulation we use a model which only includes the radiative evolution of the radio galaxy (\citealp{komissarov1994}). In particular, the radiative evolution is dominated by the synchrotron emission with a contribution from the inverse Compton scattering whose significance increases with redshift. This means that no dynamical evolution is considered, i.e. no evolution of the magnetic field nor of the volume with time. This assumption is usually made when calculating the age of the radio plasma using spectral ageing models (continuous injection, CI \citealp{jaffe1973}; and CI-off, \citealp{komissarov1994}, \citealp{murgia2011}) and therefore it is relevant to consider  here.

The parameters used in the simulation are reported in Column 2 of Table \ref{tab:param}. The magnetic field energy density is assumed to have an average fixed value of B=5 $\mu G$ for all sources, which is equal to the median equipartition value found for the sources in the B2 catalogue (\citealp{parma1999}).

\textbf{\subsubsection{Simulation with radiative and dynamical evolution}}
\label{simulationsfull}

Dynamical processes must also play a substantial role in the radio galaxy evolution. For this reason we expand the radiative model described in Section 4.2.1 to include a convenient dynamical model of the source evolution suitable for low-power radio galaxies. The model reproduces two distinct phases of expansion to describe both the active phase and inactive phase of the radio source.

In the active phase we assume that the expansion of the source is driven by the jet, according to the pressure-limiting expansion model presented in \cite{luo2010}. This considers that as the source expands the internal lobe pressure  continuously finds a balance with the external pressure that decreases outwardly. This is assumed to happen on a timescale much shorter than the age of the source, so that we can always consider the lobe pressure $\rm p_{lobe}$ to be equal to the ambient pressure $\rm p_{amb}$. In the inactive phase we assume instead that the lobes rise buoyantly through the hot atmosphere at a  velocity equal to half the local sound speed $\rm c_s$ (\citealp{ensslin2002}).  As a result, the lobe pressure decreases with time, and the source expands adiabatically and isotropically. The second phase is called the ``bubble phase". The transition between the two phases occurs at the the time when the speed of the jet-driven expansion (derived from equation 4 in \citealp{luo2010}) equals the constant speed of the bubble phase. As a first approximation, we do not include here multiple episodes of jet activity. 

We assume that the lobes expand in an external environment, i.e. the halo of its parent galaxy, whose density profile  $\rho_m$  scales with radial distance $r$ following a power law with power $\beta$:

\begin{equation}
\rho_m \propto r^{-\beta}
\end{equation}

\noindent In particular, the volume of the radio galaxy evolves with time following a piece-wise power law whose power depends on the surrounding environment and on the source age as

\begin{equation}
\rm V(t) \propto \begin{cases} t ^{3/(3-\beta)}, & if \ \ \ \ t<t_{on} \ \ \ \ \rm (Luo \ \& \ Sadler \ 2010)\\
t ^{\beta/ \Gamma}, & if \ \ \ \ t>t_{on} \ \ \ \ \rm (adiabatic \ expansion)\end{cases}
,\end{equation}

\noindent where t is the source age, $t_{on}$ is the length of its active phase, and $\rm \Gamma=4/3$ is the adiabatic index. The proportionality is justified because the model does not depend on the absolute volume value, but only on its variations over time. 

Similarly, the magnetic energy density $B$, which is assumed to be coupled with the plasma, evolves with time following a piece-wise power law whose power depends on the surrounding environment and on the source age as

\begin{equation}
\rm B(t) = \begin{cases}B_{0}t ^{-2\beta/(3-\beta)}, & if \ \ \ \ t<t_{on}\ \ \ \ \rm (Luo \ \& \ Sadler \ 2010)\\
B_{0}t ^{-\beta/2}, & if \ \ \ \ t>t_{on} \ \ \ \ \rm (adiabatic \ expansion)\end{cases}
,\end{equation}

\noindent where $B_{0}$ is the magnetic energy density value at the initial time $\rm t_0$. We describe the derivation of $B_0$ in detail below.

The parameters used in this model are listed in Table \ref{tab:param}. The values and distributions of the parameters in common with the previous model are kept equal. In the following sections we discuss in more detail the parameters that are added in this model.

\bigskip

\subsubsection*{\textbf{Radial density profile exponent  \boldmath $\beta$ \unboldmath}}
\label{beta}

The ambient medium in which the lobes expand is assumed to have a power-law radial density profile $\rm \rho_m(r),$ as discussed in Sec. \ref{simulationsfull}. As a first-order approximation, we keep the value of the power exponent $\beta$ fixed to 1.5 for all galaxies (\citealp{luo2010}, \citealp{morganti1988}).

\subsubsection*{\textbf{Initial magnetic energy density \boldmath $\rm B_0$} \unboldmath}
\label{b0}

The initial magnetic energy density $B_0$ is one of the main parameters of the model  which dictates most of the evolution of the
radio spectrum, and should be carefully normalized. Unfortunately, despite our understanding of FRII radio galaxies (e.g. \citealp{croston2005}, \citealp{harwood2016}), magnetic fields in low-power radio galaxies are still not well constrained observationally. We therefore need to rely on the equipartition condition to derive the normalization of the magnetic energy density $B_0$. We assume here that the electron-to-proton ratio is  k=1 as suggested by \cite{croston2008} for lobed FRIs.

Here we assume that at an initial time, $\rm t_0=0.1 \ Myr$, each source has a size which depends on its jet power and the ambient medium according to equation (4) of \cite{luo2010}. As we are using a pressure-limiting expansion with $\rm p_{lobe} \sim p_{amb}$, we can derive the internal lobe pressure as 

\begin{equation}
p_{lobe,0}=p_{0}(r_{0}/r)^{\beta}
,\end{equation}

\noindent where $p_{0}$ and $r_{0}$ are used to normalize the power law, and are assumed to have the following values $p_{0}=3\times10^{11}$ Pa and $r_{0}=$2  kpc (\citealp{luo2010}). Since we assume here the density profile $\rm \rho_m(r)$ to be a power law, these values just serve as normalization for the power law.  
Assuming equipartition between magnetic field energy and particle energy we can then derive

\begin{equation}
B_0 = (p_{lobe,0}\times \mu_0)^{0.5}
,\end{equation}

\noindent where  $\mu_0$ is the magnetic constant.

According to this derivation, at the time $t_0$ the lobes of low jet power sources are smaller and therefore more confined to the higher density regions of the ambient medium. Since the pressure of the external gas is higher and we  assume a balance between the lobe pressure and the external pressure, these sources have higher lobe pressure and therefore higher equipartition magnetic fields.

\begin{figure*}[h!]
\centering
\includegraphics[width=1\textwidth]{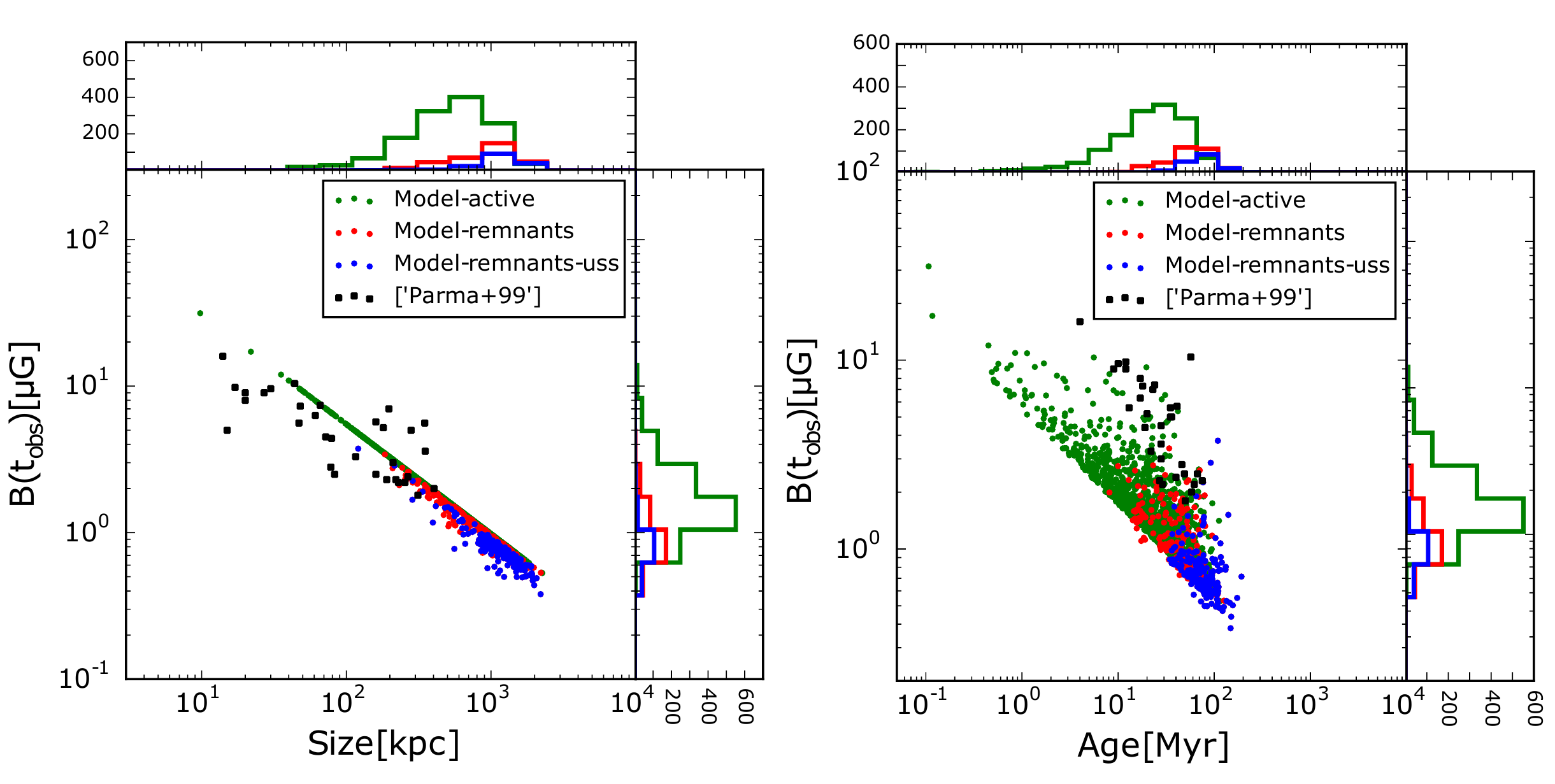}
\caption{Magnetic field as a function of age (right panel) and source size (left panel). Green, red, and blue circles represent active, remnant, and ultra-steep remnant sources, respectively, as predicted by the model with radiative and dynamical evolution (Section 4.3.2). Black squares represent sources of the B2 sample (\citealp{parma1999}).  }
\label{fig:parma}       
\end{figure*}

To confirm the robustness of this normalization we have compared the magnetic fields derived by \cite{parma1999} for the B2 sample to those obtained in our mock catalogues at $t_{obs}$ (Section 4.3.2). In particular, in Figure \ref{fig:parma} we show how the magnetic field correlates with age and size for sources in the mock catalogue and for sources in the B2 sample. The values obtained from the simulations are well matched by those observed in the B2 sample.

\subsubsection*{\textbf{Sound speed  \boldmath $c_s$} \unboldmath }
\label{cs}

We assume that the expansion in the second dynamical phase (the remnant phase) happens at a constant velocity equal to 500 $\rm km\ s^{-1}$. This value corresponds to half of the local sound speed $\rm c_s$ for an intergalactic medium of intermediate density (\citealp{ensslin2002}). Lower values cause the sources with low jet powers ($\rm Q_{jet}\lesssim5\times10^{35}$) to enter the bubble phase when the jets are still active. In these cases, the jets are so weak that the isotropic adiabatic expansion starts to dominate at earlier stages. 

Typical values of the sound speeds in the intergalactic medium vary in the range $\rm \sim500-1500 \ km \ s^{-1}$ with rich cluster environments having higher $c_s$ with respect to small groups.

\subsection{Simulation results }

\begin{figure*}
\centering
\includegraphics[width=1.0\textwidth]{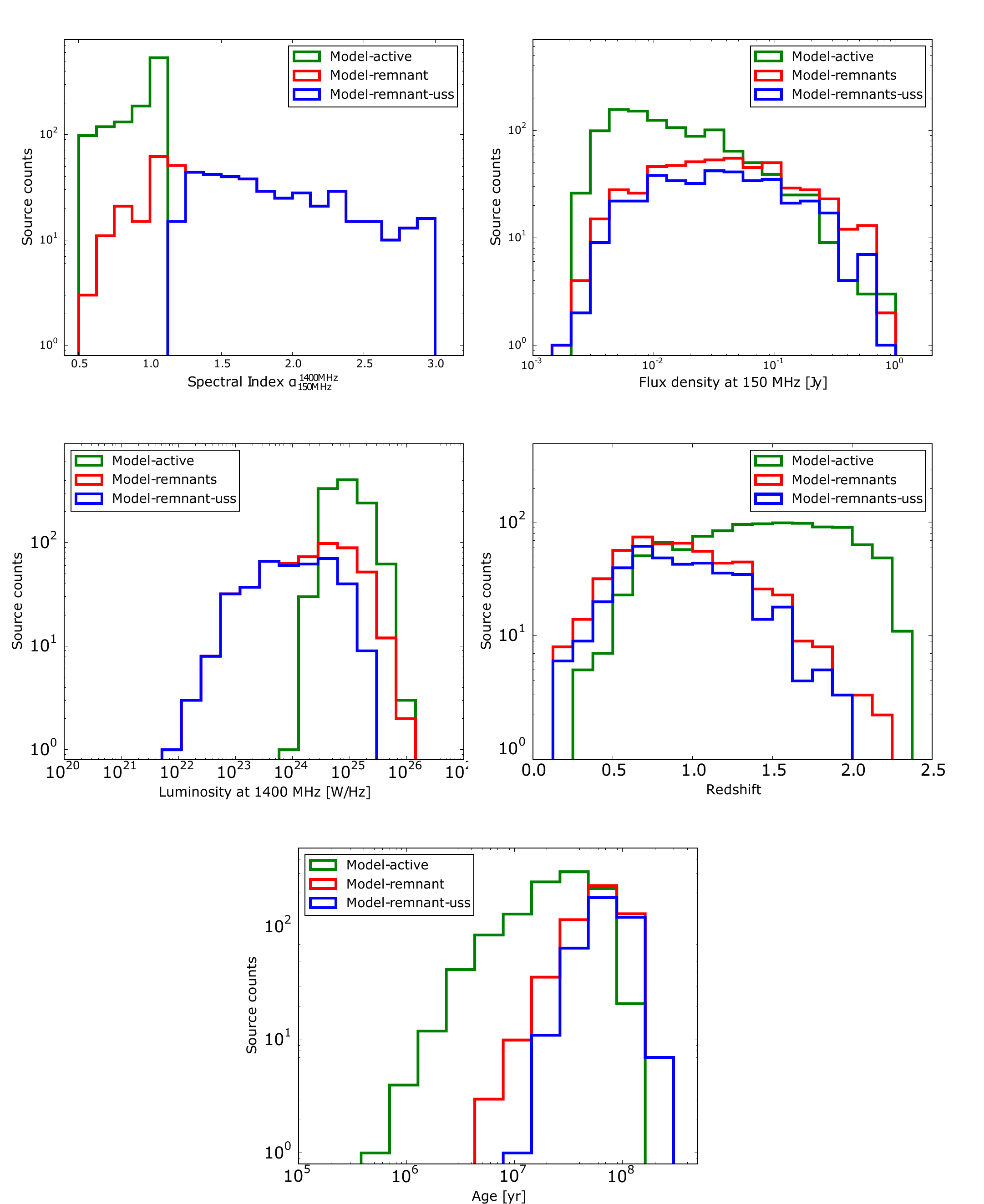}
\caption{Main output parameters for the mock catalogue produced assuming radiative evolution models only, as described in Section 4.2.1. The green line represents active sources, the red line represents remnant sources, the blue line represents ultra-steep spectrum remnant sources with $\rm \alpha_{150}^{1400}>1.2$.}
\label{fig:plotscioff}  
\end{figure*}

\begin{figure*}
\centering
\includegraphics[width=1.0\textwidth]{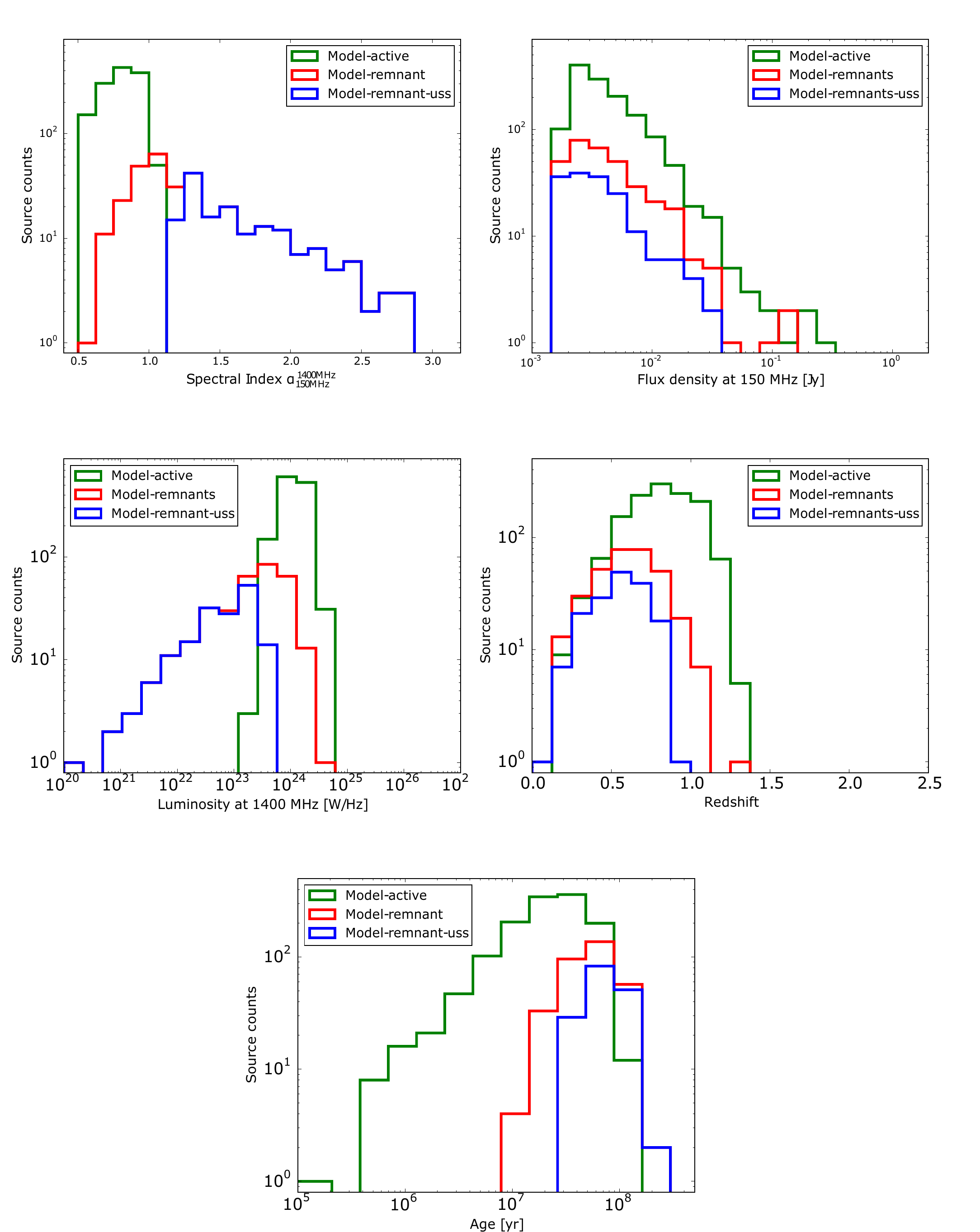}
\caption{Main output parameters for the mock catalogue produced assuming radiative and dynamical evolution models, as described in Section 4.2.2. The green line represents active sources, the red line represents remnant sources, the blue line represents ultra-steep spectrum remnant sources with $\rm \alpha_{150}^{1400}>1.2$.}
\label{fig:plots}
\end{figure*}

In this section we describe the main results of the simulations presented in the former sections. Table \ref{tab:mockcatalogues} shows the fractions of different sources found in the two simulations. Figures \ref{fig:plotscioff} and \ref{fig:plots} show the main distributions of the output parameters for the two simulations for active ($t_{obs}<t_{on}$), remnant ($t_{obs}>t_{on}$), and ultra-steep spectrum remnant ($t_{obs}<t_{on}$ and $\alpha_{150}^{1400}>1.2$) sources.

\begin{small}
\begin{table}[h]
        \centering
        \caption{Results of Monte Carlo simulations.  }
        \small
                \begin{tabular}{*3c}
                \hline
                \hline
            Number of sources & Radiative & Radiative + Dynamic  \\
            in the sample  &  evolution & evolution\\
                ($\rm S_{150MHz}>1.5 mJy$) & & \\               
                \hline
                \hline
                Total* & 1609 & 1665\\ 
                Active ($\rm t_{obs}<t_{on}$) & 1073 (66\%) & 1317 (79\%)\\
            Remnants ($\rm t_{obs}>t_{on}$)&  536 (33\%)&  329 (20\%)\\
                Ultra-steep spectrum & 387 (24\%) & 165 (10\%)\\
                ($\rm t_{obs}<t_{on}$ and $\rm \alpha_{150}^{1400}>1.2$) &&\\
                Ultra-steep spectrum & 444 (28\%) & 321 (19\%)\\
                ($\rm t_{obs}<t_{on}$ and $\rm \alpha_{150}^{5000}>1.2$) &&\\
                \hline
                \hline  
                \end{tabular}
                \begin{tablenotes}
        \small
        \item \textbf{Notes.} $^*$Sources selected above a flux density of 1.5 mJy and produced from a simulation of 1.5$\rm \times 10^5$ sources.
        \end{tablenotes}
          \label{tab:mockcatalogues}
\end{table}
\end{small}

\subsubsection{Results of the simulation with radiative evolution model}
\label{resultsrad}

In this section we discuss the results of the simulation with the  radiative evolution model presented in Section 4.2.1. 
As summarized in Table \ref{tab:mockcatalogues}, the mock catalogue contains a total of 1609 sources, of which 1073 (66\%) are active ($t_{obs}<t_{on}$) and 536 (33\%) are remnants ($t_{obs}>t_{on}$). In particular, 387 sources (24\% of the entire catalogue) have ultra-steep spectra in the range 150-1400 MHz with $\rm \alpha_{150}^{1400}>1.2$.

The major result of this simulation is that the fraction of ultra-steep spectrum FRI remnant radio galaxies (24\% of the catalogue) is very hard to reconcile with the upper limit of <15\% that we derived from the empirical catalogue (see Section \ref{skad}).
This shows that models which are classically used to derive radiative ages of FRIs give inconsistent results in terms of the fraction of remnants observed in flux limited samples. This supports the finding of \cite{harwood2017} that modelling of integrated radio spectra with available models, i.e. CI (\citealp{jaffe1973}) and CI-off (\citealp{komissarov1994}), possibly leads to unrealistic source ages and  shows the limitation of this approach.

In Figure \ref{fig:plotscioff} we show the main distributions of the output parameters for the simulation with radiative evolution alone. 

\subsubsection{Results of  the simulation with radiative and dynamical evolution models}
\label{resultsfull}

In this section we describe the results obtained from the simulation with both radiative and dynamical evolution models presented in Section 4.2.2. As summarized in Table \ref{tab:mockcatalogues}, the mock catalogue contains a total of 1665 sources, of which 1317 (79\%) are active ($t_{obs}<t_{on}$) and 329 (20\%) are remnants ($t_{obs}>t_{on}$). Interestingly, only half of the remnant sources (165 sources and 10\% of the entire catalogue) have ultra-steep spectra in the range 150-1400 MHz with $\rm \alpha_{150}^{1400}>1.2$. 

These results confirm that remnant radio galaxies with ultra-steep spectra in the range 150-1400 MHz represent only a fraction of the entire remnant population, as was demonstrated by \cite{godfrey2017} for FRII sources. The results also validate our decision to select remnant radio galaxies in the Lockman Hole field using three complementary selection criteria as described in Section \ref{selection}. The percentage of ultra-steep spectra remnant radio galaxies found in this simulation including dynamical evolution in the remnant phase is 10\%, consistent with the observed upper limit for FRIs (<15\%; see  Section \ref{skad}). Given the large number of parameters in the model and the first-order nature of the calculations, the agreement between the two values is impressive and contrasts with the previous model for which no dynamical evolution was implemented in the remnant phase. These results demonstrate that dynamical evolution of FRI radio galaxies continues in the remnant phase.

More information from the simulation results can be obtained by looking at Figure \ref{fig:plots}. The top left panel of Figure 6 shows the spectral index distribution of the sources in the mock catalogue. The distribution has a median value of $\rm \alpha$=0.78 in agreement with observations and has a tail that extends to higher values. In accordance with the CI model, active sources have values in the range  0.5-1, while remnant sources dominate the ultra-steep spectrum tail. From this plot it is also clear that not all remnant radio galaxies show ultra-steep spectra. 

The flux density distribution at 150 MHz (Figure \ref{fig:plots}, top right) shows the expected decrease in the number of sources at high flux densities. The luminosity distribution at 1400 MHz (Figure \ref{fig:plots},  centre left) peaks around $\rm 10^{24} \ WHz^{-1}$ in agreement with observed low-power radio galaxies, with a tail of remnant sources that extends to $\rm 10^{20} \ WHz^{-1}$. Figure \ref{fig:plots}, centre right, demonstrates that low-power radio sources in such a flux limited sample ($\rm S_{150}$>1.5 mJy) are not expected to be located at redshifts much higher than 1.3. In particular, due to their faster luminosity evolution and the inverse Compton scattering at high redshifts, remnant radio galaxies are mostly observed at redshifts of less than 1. Figure \ref{fig:plots}, bottom, shows that after the switch off, the remnant plasma quickly becomes invisible. About 70\% of the remnant sources have ages <1.5 $\rm \times \ t_{on}$. This suggests that the luminosity evolution of the plasma after the jets switch off is very fast, due to adiabatic and radiative losses and to the decreasing magnetic field. In particular, ultra-steep remnants occupy the oldest tail of the distribution.

\subsection{Discussion of the Monte Carlo simulation results}

The simulations presented in the previous sections are based on simplified models of radio galaxy evolution and, in this particular paper, we focus on one specific source type, i.e. lobed FRIs. Here we briefly comment the uncertainties connected with the choice of parameters and the implications that they have on the robustness of the results. Among all the parameters of the simulations, we have identified three that appear to have the most influence on  the final results. The first  is the initial magnetic energy density $B_0$ which has been derived according to equipartition conditions. Various studies have shown, at least for most FRII radio galaxies, that this approximation is not correct and an average correction factor of 0.7 should be applied to equipartition magnetic field (\citealp{croston2005}, \citealp{hardcastle2016}). Although no such studies have been performed on FRI radio galaxies, we cannot exclude that variations from the equipartition conditions are present, as observed for higher power sources. In particular, lower values of the magnetic energy density would increase the remnant fraction, while higher values would decrease the remnant fraction. As a control test we ran the same simulation decreasing B by a factor 0.7 and  found that the fraction of ultra-steep remnants increases from 10\% to 11\%.

The second parameter that plays a major role in the final result is the slope of the radial density profile of the external medium $\beta$. We investigate how increasing $\beta$ to 1.8 affects the resulting remnant fraction and we find a decrease of about a factor 2. A value of $\beta$=1.8 also provides a worse match with the trends observed in the B2 catalogue, which are  plotted in Fig. \ref{fig:parma}. 

Another parameter that is worth discussing is the source active time $t_{on}$. At present, our knowledge of the radio galaxy duty cycle is limited to very few single-object studies, and information on its distribution among the radio galaxy population is not available (e.g. \citealp{konar2013, shulevski2017}). We have explored the variation in the results when modifying the parameters of the distribution of $t_{on}$, and we find that the fraction of ultra-steep remnant can go as low as 6\% when increasing  $\overline{t_{on}}$ to 60 Myr and increasing  $\sigma_{t_{on}}$ to 100 Myr. Likely, we will be soon able to put firmer constraints as bigger samples of radio galaxies are characterized by new radio facilities like LOFAR. 

Despite the uncertainties discussed above, the results of the simulation with radiative and dynamical models is promising and  demonstrate that by including a dynamical component in the spectral models in addition to the radiative component, we can obtain agreement between the fraction of remnant sources in the simulated catalogue and the observed upper limit.

Finally, it is interesting to note that simulations predict that, by including 5000-MHz data in the selection with the ultra-steep spectrum criterion, it would be possible to recover almost the entire remnant population (see Table \ref{tab:mockcatalogues}). The need to include frequencies > 1400 MHz in the ultra-steep spectrum selection was already recognized by \cite{brienza2016}. The validity of this prediction could be further tested by including data at frequencies >1400 MHz in the empirical source selection. However, such high-frequency observations at such low flux limits are currently not available over a large fraction of the sky.

\section{Summary and conclusions}

In this paper we have presented a search for remnant radio galaxies in the extra-galactic Lockman-Hole field in order to assess their fraction in low-frequency radio samples. Using both spectral and morphological criteria, we have selected 23 remnant radio galaxy candidates which will be confirmed with dedicated future observations. Furthermore, we have created mock catalogues of low-power radio galaxies using Monte Carlo simulations to investigate whether current radiative and dynamical models of the radio galaxy evolution predict a fraction of remnant radio galaxies which is consistent with observations. The main findings are as follows:
\\

(i) Ultra-steep spectrum remnant sources with $\rm \alpha_{150}^{1400}$>1.2 represent a fraction of <4.1\% [3.7\%-6.3\%] of the entire Lockman Hole low resolution catalogue and <6.6\%  [4.3\%-9.2\%] of the Lockman-WSRT catalogue. In particular, we expect the fraction of lobed FRIs to be $\sim 43$\% of the entire radio catalogue and can  therefore put an upper limit on the fraction of remnant lobed FRIs with ultra-steep spectrum in the Lockman-WSRT of < 0.066/0.43 $\sim$15\% (in the range [10\%-21\%]). Simulations based on radiative and dynamical models predict a remnant fraction  of 10\%, which is consistent with observations. On the contrary, simulations based on radiative models only overpredict the number of ultra-steep spectrum remnants by almost a factor of 2. This demonstrates that dynamical evolution plays an important role in the remnant phase of lobed FRI radio galaxies. Most importantly, it shows that models which are classically used to derive radiative ages of FRIs give inconsistent results in terms of the fraction of remnants observed in flux limited samples. By neglecting adiabatic cooling, and magnetic field evolution during the remnant phase, radiative ages overestimate the remnant age, and therefore can be considered as upper limits. 
\\

(ii) Mock catalogues show that ultra-steep spectrum remnant sources with $\rm \alpha_{150}^{1400}$>1.2 represent only a fraction of the entire FRIs remnant radio galaxy population ($\sim$50\%) and, in particular, represent the oldest tail of the age distribution. When including 5000 MHz observations in the selection the fraction of remnants is almost entirely recovered. This demonstrates the need to include  frequencies >1400 MHz or additional selection methods in order to collect the entire population. This is in agreement with the findings by \cite{godfrey2017} for FRII radio galaxies.
\\

(iii) Simulations predict a very rapid drop in remnant radio galaxies above 100 Myr. About 70\% of the remnant sources have ages <1.5 $\rm \times \ t_{on}$. This suggest that the luminosity evolution of the plasma after the jets switch off is very fast, due to adiabatic and radiative losses and to the decreasing magnetic field. 
\\

(iv) Morphology and radio core prominence can be used as complementary selection criteria to identify remnant radio galaxies, thus avoiding any age bias. A fraction of <46\% of morphologically selected remnants and a fraction of $\sim$10\% of core prominence selected remnants are found to have ultra-steep spectra with $\rm \alpha_{150}^{1400}$>1.2. This result is in agreement with the simulation results showing that ultra-steep spectrum remnants represent only a fraction of the entire population and that remnant plasma undergoes a very rapid luminosity evolution.
\\

(v) The spectral curvature criterion in the frequency range 150-1400 MHz is not ideal for selecting remnant radio galaxies. Because the steepening of the spectrum at early stages occurs mostly at higher frequency, including 5000-MHz observations in this kind of analysis is essential for the selection to work. Unfortunately, observations at such low flux limit at 5000 MHz is currently not available over a large fraction of the sky. 
\\

\begin{acknowledgements}
The research leading to these results has received funding from the European Research Council under the European Union's Seventh Framework Programme (FP/2007-2013) / ERC Advanced Grant RADIOLIFE-320745 PI R. Morganti. LOFAR, the Low Frequency Array designed and constructed by ASTRON (Netherlands Institute for Radio Astronomy), has facilities in several countries that are owned by various parties (each with its own funding sources), and that are collectively operated by the International LOFAR Telescope (ILT) foundation under a joint scientific policy. This research has made use of the NASA/IPAC Extragalactic Database (NED), which is operated by the Jet Propulsion Laboratory, California Institute of Technology, under contract with the National Aeronautics and Space Administration. This research made use of APLpy, an open-source plotting package for Python hosted at http://aplpy.github.com. EKM acknowledges support from the Australian Research Council Centre of Excellence for All-sky Astrophysics (CAASTRO), through project number CE110001020. MJH acknowledges support from the UK Science and Technology Facilities Council [ST/M001008/1]. TS acknowledges support from the ERC Advanced Investigator programme NewClusters 321271.
      
\end{acknowledgements}

\bibliographystyle{aa}
\bibliography{brienza_remnants_august17.bib}

\begin{thebibliography}{76}
\expandafter\ifx\csname natexlab\endcsname\relax\def\natexlab#1{#1}\fi

\bibitem[{{Afonso} {et~al.}(2011){Afonso}, {Bizzocchi}, {Ibar}, {Grossi},
  {Simpson}, {Chapman}, {Jarvis}, {R{\"o}ttgering}, {Norris}, {Dunlop},
  {Ivison}, {Messias}, {Pforr}, {Vaccari}, {Seymour}, {Best},
  {Gonz{\'a}lez-Solares}, {Farrah}, {Fernandes}, {Huang}, {Lacy}, {Maraston},
  {Marchetti}, {Mauduit}, {Oliver}, {Rigopoulou}, {Stanford}, {Surace}, \&
  {Zeimann}}]{afonso2011}
{Afonso}, J., {Bizzocchi}, L., {Ibar}, E., {et~al.} 2011, \apj, 743, 122

\bibitem[{{AMI Consortium} {et~al.}(2011){AMI Consortium}, {Davies}, {Franzen},
  {Waldram}, {Grainge}, {Hobson}, {Hurley-Walker}, {Lasenby}, {Olamaie},
  {Pooley}, {Riley}, {Rodr{\'{\i}}guez-Gonz{\'a}lvez}, {Saunders}, {Scaife},
  {Schammel}, {Scott}, {Shimwell}, {Titterington}, \& {Zwart}}]{davies2011}
{AMI Consortium}, {Davies}, M.~L., {Franzen}, T.~M.~O., {et~al.} 2011, \mnras,
  415, 2708

\bibitem[{{Becker} {et~al.}(1995){Becker}, {White}, \& {Helfand}}]{becker1995}
{Becker}, R.~H., {White}, R.~L., \& {Helfand}, D.~J. 1995, \apj, 450, 559

\bibitem[{{Blandford} \& {Ostriker}(1978)}]{blandford1978}
{Blandford}, R.~D. \& {Ostriker}, J.~P. 1978, \apjl, 221, L29

\bibitem[{{Blundell} {et~al.}(1999){Blundell}, {Rawlings}, \&
  {Willott}}]{blundell1999}
{Blundell}, K.~M., {Rawlings}, S., \& {Willott}, C.~J. 1999, \aj, 117, 677

\bibitem[{{Brienza} {et~al.}(2016){Brienza}, {Godfrey}, {Morganti}, {Vilchez},
  {Maddox}, {Murgia}, {Orru}, {Shulevski}, {Best}, {Br{\"u}ggen}, {Harwood},
  {Jamrozy}, {Jarvis}, {Mahony}, {McKean}, \& {R{\"o}ttgering}}]{brienza2016}
{Brienza}, M., {Godfrey}, L., {Morganti}, R., {et~al.} 2016, \aap, 585, A29

\bibitem[{{Carilli} {et~al.}(1991){Carilli}, {Perley}, {Dreher}, \&
  {Leahy}}]{carilli1991}
{Carilli}, C.~L., {Perley}, R.~A., {Dreher}, J.~W., \& {Leahy}, J.~P. 1991,
  \apj, 383, 554

\bibitem[{{Cohen} {et~al.}(2007){Cohen}, {Lane}, {Cotton}, {Kassim}, {Lazio},
  {Perley}, {Condon}, \& {Erickson}}]{cohen2007}
{Cohen}, A.~S., {Lane}, W.~M., {Cotton}, W.~D., {et~al.} 2007, \aj, 134, 1245

\bibitem[{{Cohen} {et~al.}(2004){Cohen}, {Lane}, {Kassim}, {Lazio}, {Cotton},
  {Perley}, {Condon}, \& {Erickson}}]{cohen2004}
{Cohen}, A.~S., {Lane}, W.~M., {Kassim}, N.~E., {et~al.} 2004, in Bulletin of
  the American Astronomical Society, Vol.~36, American Astronomical Society
  Meeting Abstracts, 1488

\bibitem[{{Colla} {et~al.}(1970){Colla}, {Fanti}, {Ficarra}, {Formiggini},
  {Gandolfi}, {Grueff}, {Lari}, {Padrielli}, {Roffi}, {Tomasi}, \&
  {Vigotti}}]{colla1970}
{Colla}, G., {Fanti}, C., {Ficarra}, A., {et~al.} 1970, \aaps, 1, 281

\bibitem[{{Condon} {et~al.}(1998){Condon}, {Cotton}, {Greisen}, {Yin},
  {Perley}, {Taylor}, \& {Broderick}}]{condon1998}
{Condon}, J.~J., {Cotton}, W.~D., {Greisen}, E.~W., {et~al.} 1998, \aj, 115,
  1693

\bibitem[{{Croston}(2008)}]{croston2008}
{Croston}, J.~H. 2008, in Astronomical Society of the Pacific Conference
  Series, Vol. 386, Extragalactic Jets: Theory and Observation from Radio to
  Gamma Ray, ed. T.~A. {Rector} \& D.~S. {De Young}, 335

\bibitem[{{Croston} {et~al.}(2005){Croston}, {Hardcastle}, {Harris}, {Belsole},
  {Birkinshaw}, \& {Worrall}}]{croston2005}
{Croston}, J.~H., {Hardcastle}, M.~J., {Harris}, D.~E., {et~al.} 2005, \apj,
  626, 733

\bibitem[{{De Breuck} {et~al.}(2000){De Breuck}, {van Breugel},
  {R{\"o}ttgering}, \& {Miley}}]{debreuck2000}
{De Breuck}, C., {van Breugel}, W., {R{\"o}ttgering}, H.~J.~A., \& {Miley}, G.
  2000, \aaps, 143, 303

\bibitem[{{de Gasperin} {et~al.}(2015){de Gasperin}, {Ogrean}, {van Weeren},
  {Dawson}, {Br{\"u}ggen}, {Bonafede}, \& {Simionescu}}]{degasperin2015}
{de Gasperin}, F., {Ogrean}, G.~A., {van Weeren}, R.~J., {et~al.} 2015, \mnras,
  448, 2197

\bibitem[{{de Ruiter} {et~al.}(1990){de Ruiter}, {Parma}, {Fanti}, \&
  {Fanti}}]{deruiter1990}
{de Ruiter}, H.~R., {Parma}, P., {Fanti}, C., \& {Fanti}, R. 1990, \aap, 227,
  351

\bibitem[{{Eilek}(1996)}]{eilek1996}
{Eilek}, J.~A. 1996, in Astronomical Society of the Pacific Conference Series,
  Vol. 100, Energy Transport in Radio Galaxies and Quasars, ed. P.~E. {Hardee},
  A.~H. {Bridle}, \& J.~A. {Zensus}, 281

\bibitem[{{En{\ss}lin} \& {Br{\"u}ggen}(2002)}]{ensslin2002}
{En{\ss}lin}, T.~A. \& {Br{\"u}ggen}, M. 2002, \mnras, 331, 1011

\bibitem[{{Fanaroff} \& {Riley}(1974)}]{fanaroff1974}
{Fanaroff}, B.~L. \& {Riley}, J.~M. 1974, \mnras, 167, 31P

\bibitem[{{Fanti} {et~al.}(1978){Fanti}, {Gioia}, {Lari}, \&
  {Ulrich}}]{fanti1978}
{Fanti}, R., {Gioia}, I., {Lari}, C., \& {Ulrich}, M.~H. 1978, \aaps, 34, 341

\bibitem[{{Feretti} {et~al.}(1984){Feretti}, {Gioia}, {Giovannini},
  {Gregorini}, \& {Padrielli}}]{feretti1984}
{Feretti}, L., {Gioia}, I.~M., {Giovannini}, G., {Gregorini}, L., \&
  {Padrielli}, L. 1984, \aap, 139, 50

\bibitem[{{Giovannini} {et~al.}(1988){Giovannini}, {Feretti}, {Gregorini}, \&
  {Parma}}]{giovannini1988}
{Giovannini}, G., {Feretti}, L., {Gregorini}, L., \& {Parma}, P. 1988, \aap,
  199, 73

\bibitem[{{Godfrey} {et~al.}(2017){Godfrey}, {Morganti}, \&
  {Brienza}}]{godfrey2017}
{Godfrey}, L.~E.~H., {Morganti}, R., \& {Brienza}, M. 2017, ArXiv e-prints
  1706.05909

\bibitem[{{Hardcastle}(2013)}]{hardcastle2013}
{Hardcastle}, M.~J. 2013, \mnras, 433, 3364

\bibitem[{{Hardcastle} {et~al.}(2016){Hardcastle}, {G{\"u}rkan}, {van Weeren},
  {Williams}, {Best}, {de Gasperin}, {Rafferty}, {Read}, {Sabater}, {Shimwell},
  {Smith}, {Tasse}, {Bourne}, {Brienza}, {Br{\"u}ggen}, {Brunetti},
  {Chy{\.z}y}, {Conway}, {Dunne}, {Eales}, {Maddox}, {Jarvis}, {Mahony},
  {Morganti}, {Prandoni}, {R{\"o}ttgering}, {Valiante}, \&
  {White}}]{hardcastle2016}
{Hardcastle}, M.~J., {G{\"u}rkan}, G., {van Weeren}, R.~J., {et~al.} 2016,
  \mnras, 462, 1910

\bibitem[{{Harwood}(2017)}]{harwood2017}
{Harwood}, J.~J. 2017, \mnras, 466, 2888

\bibitem[{{Harwood} {et~al.}(2016){Harwood}, {Croston}, {Intema}, {Stewart},
  {Ineson}, {Hardcastle}, {Godfrey}, {Best}, {Brienza}, {Heesen}, {Mahony},
  {Morganti}, {Murgia}, {Orr{\'u}}, {R{\"o}ttgering}, {Shulevski}, \&
  {Wise}}]{harwood2016}
{Harwood}, J.~J., {Croston}, J.~H., {Intema}, H.~T., {et~al.} 2016, \mnras,
  458, 4443

\bibitem[{{Harwood} {et~al.}(2015){Harwood}, {Hardcastle}, \&
  {Croston}}]{harwood2015}
{Harwood}, J.~J., {Hardcastle}, M.~J., \& {Croston}, J.~H. 2015, \mnras, 454,
  3403

\bibitem[{{Harwood} {et~al.}(2013){Harwood}, {Hardcastle}, {Croston}, \&
  {Goodger}}]{harwood2013}
{Harwood}, J.~J., {Hardcastle}, M.~J., {Croston}, J.~H., \& {Goodger}, J.~L.
  2013, \mnras, 435, 3353

\bibitem[{{Hogg}(1999)}]{hogg1999}
{Hogg}, D.~W. 1999, ArXiv Astrophysics e-prints 9905116

\bibitem[{{Ibar} {et~al.}(2009){Ibar}, {Ivison}, {Biggs}, {Lal}, {Best}, \&
  {Green}}]{ibar2009}
{Ibar}, E., {Ivison}, R.~J., {Biggs}, A.~D., {et~al.} 2009, \mnras, 397, 281

\bibitem[{{Intema} {et~al.}(2017){Intema}, {Jagannathan}, {Mooley}, \&
  {Frail}}]{intema2017}
{Intema}, H.~T., {Jagannathan}, P., {Mooley}, K.~P., \& {Frail}, D.~A. 2017,
  \aap, 598, A78

\bibitem[{{Jaffe} \& {Perola}(1973)}]{jaffe1973}
{Jaffe}, W.~J. \& {Perola}, G.~C. 1973, \aap, 26, 423

\bibitem[{{Kaiser}(2009)}]{kaiser2009}
{Kaiser}, C.~R. 2009, Astronomische Nachrichten, 330, 270

\bibitem[{{Kaiser} \& {Best}(2007)}]{kaiser2007}
{Kaiser}, C.~R. \& {Best}, P.~N. 2007, \mnras, 381, 1548

\bibitem[{{Kapinska} {et~al.}(2015){Kapinska}, {Hardcastle}, {Jackson}, {An},
  {Baan}, \& {Jarvis}}]{kapinska2015}
{Kapinska}, A.~D., {Hardcastle}, M., {Jackson}, C., {et~al.} 2015, Advancing
  Astrophysics with the Square Kilometre Array (AASKA14), 173

\bibitem[{{Kardashev}(1962)}]{kardashev1962}
{Kardashev}, N.~S. 1962, \sovast, 6, 317

\bibitem[{{Ker} {et~al.}(2012){Ker}, {Best}, {Rigby}, {R{\"o}ttgering}, \&
  {Gendre}}]{ker2012}
{Ker}, L.~M., {Best}, P.~N., {Rigby}, E.~E., {R{\"o}ttgering}, H.~J.~A., \&
  {Gendre}, M.~A. 2012, \mnras, 420, 2644

\bibitem[{{Komissarov} \& {Gubanov}(1994)}]{komissarov1994}
{Komissarov}, S.~S. \& {Gubanov}, A.~G. 1994, \aap, 285, 27

\bibitem[{{Konar} {et~al.}(2013){Konar}, {Hardcastle}, {Jamrozy}, \&
  {Croston}}]{konar2013}
{Konar}, C., {Hardcastle}, M.~J., {Jamrozy}, M., \& {Croston}, J.~H. 2013,
  \mnras, 430, 2137

\bibitem[{{Laing} \& {Bridle}(2013)}]{laing2013}
{Laing}, R.~A. \& {Bridle}, A.~H. 2013, \mnras, 432, 1114

\bibitem[{{Luo} \& {Sadler}(2010)}]{luo2010}
{Luo}, Q. \& {Sadler}, E.~M. 2010, \apj, 713, 398

\bibitem[{{Mahony} {et~al.}(2016){Mahony}, {Morganti}, {Prandoni}, {van
  Bemmel}, {Shimwell}, {Brienza}, {Best}, {Br{\"u}ggen}, {Calistro Rivera}, {de
  Gasperin}, {Hardcastle}, {Harwood}, {Heald}, {Jarvis}, {Mandal}, {Miley},
  {Retana-Montenegro}, {R{\"o}ttgering}, {Sabater}, {Tasse}, {van Velzen}, {van
  Weeren}, {Williams}, \& {White}}]{mahony2016}
{Mahony}, E.~K., {Morganti}, R., {Prandoni}, I., {et~al.} 2016, \mnras, 463,
  2997

\bibitem[{{Mohan} \& {Rafferty}(2015)}]{mohan2015}
{Mohan}, N. \& {Rafferty}, D. 2015, {PyBDSM: Python Blob Detection and Source
  Measurement}, Astrophysics Source Code Library, eprint 1502.007

\bibitem[{{Morganti} {et~al.}(1988){Morganti}, {Fanti}, {Gioia}, {Harris},
  {Parma}, \& {de Ruiter}}]{morganti1988}
{Morganti}, R., {Fanti}, R., {Gioia}, I.~M., {et~al.} 1988, \aap, 189, 11

\bibitem[{{Mullin} {et~al.}(2008){Mullin}, {Riley}, \&
  {Hardcastle}}]{mullin2008}
{Mullin}, L.~M., {Riley}, J.~M., \& {Hardcastle}, M.~J. 2008, \mnras, 390, 595

\bibitem[{{Murgia} {et~al.}(2011){Murgia}, {Parma}, {Mack}, {de Ruiter},
  {Fanti}, {Govoni}, {Tarchi}, {Giacintucci}, \& {Markevitch}}]{murgia2011}
{Murgia}, M., {Parma}, P., {Mack}, K.-H., {et~al.} 2011, \aap, 526, A148

\bibitem[{{Myers} {et~al.}(2014){Myers}, {Baum}, \& {Chandler}}]{myers2014}
{Myers}, S.~T., {Baum}, S.~A., \& {Chandler}, C.~J. 2014, in American
  Astronomical Society Meeting Abstracts, Vol. 223, American Astronomical
  Society Meeting Abstracts \#223, 236.01

\bibitem[{{Oosterloo} {et~al.}(2009){Oosterloo}, {Verheijen}, {van Cappellen},
  {Bakker}, {Heald}, \& {Ivashina}}]{oosterloo2009}
{Oosterloo}, T., {Verheijen}, M.~A.~W., {van Cappellen}, W., {et~al.} 2009, in
  Wide Field Astronomy Technology for the Square Kilometre Array, 70

\bibitem[{{Pacholczyk}(1970)}]{pacholczyc1970}
{Pacholczyk}, A.~G. 1970, {Radio astrophysics. Nonthermal processes in galactic
  and extragalactic sources, Series of Books in Astronomy and Astrophysics, San
  Francisco: Freeman, 1970}

\bibitem[{{Parma} {et~al.}(1996){Parma}, {de Ruiter}, \& {Fanti}}]{parma1996}
{Parma}, P., {de Ruiter}, H.~R., \& {Fanti}, R. 1996, in IAU Symposium, Vol.
  175, Extragalactic Radio Sources, ed. R.~D. {Ekers}, C.~{Fanti}, \&
  L.~{Padrielli}, 137

\bibitem[{{Parma} {et~al.}(2007){Parma}, {Murgia}, {de Ruiter}, {Fanti},
  {Mack}, \& {Govoni}}]{parma2007}
{Parma}, P., {Murgia}, M., {de Ruiter}, H.~R., {et~al.} 2007, \aap, 470, 875

\bibitem[{{Parma} {et~al.}(1999){Parma}, {Murgia}, {Morganti}, {Capetti}, {de
  Ruiter}, \& {Fanti}}]{parma1999}
{Parma}, P., {Murgia}, M., {Morganti}, R., {et~al.} 1999, \aap, 344, 7

\bibitem[{{Prandoni} {et~al.}(in preparation){Prandoni}, {Guglielmino}, \&
  {Morganti}}]{prandoni2017}
{Prandoni}, I., {Guglielmino}, G., \& {Morganti}, R. in preparation

\bibitem[{{Rengelink} {et~al.}(1997){Rengelink}, {Tang}, {de Bruyn}, {Miley},
  {Bremer}, {R{\"o}ttgering}, \& {Bremer}}]{rengelink1997}
{Rengelink}, R.~B., {Tang}, Y., {de Bruyn}, A.~G., {et~al.} 1997, \aaps, 124,
  259

\bibitem[{{Roger} {et~al.}(1973){Roger}, {Costain}, \& {Bridle}}]{roger1973}
{Roger}, R.~S., {Costain}, C.~H., \& {Bridle}, A.~H. 1973, \aj, 78, 1030

\bibitem[{{Rottgering} {et~al.}(2006){Rottgering}, {Braun}, {Barthel}, {van
  Haarlem}, {Miley}, {Morganti}, {Snellen}, {Falcke}, {de Bruyn}, {Stappers},
  {Boland}, {Butcher}, {de Geus}, {Koopmans}, {Fender}, {Kuijpers},
  {Schilizzi}, {Vogt}, {Wijers}, {Wise}, {Brouw}, {Hamaker}, {Noordam},
  {Oosterloo}, {Bahren}, {Brentjens}, {Wijnholds}, {Bregman}, {van Cappellen},
  {Gunst}, {Kant}, {Reitsma}, {van der Schaaf}, \& {de Vos}}]{rottgering2006}
{Rottgering}, H.~J.~A., {Braun}, R., {Barthel}, P.~D., {et~al.} 2006, ArXiv
  Astrophysics e-prints

\bibitem[{{R{\"o}ttgering} {et~al.}(1994){R{\"o}ttgering}, {Lacy}, {Miley},
  {Chambers}, \& {Saunders}}]{roettgering1994}
{R{\"o}ttgering}, H.~J.~A., {Lacy}, M., {Miley}, G.~K., {Chambers}, K.~C., \&
  {Saunders}, R. 1994, \aaps, 108

\bibitem[{{Saripalli} {et~al.}(2012){Saripalli}, {Subrahmanyan}, {Thorat},
  {Ekers}, {Hunstead}, {Johnston}, \& {Sadler}}]{saripalli2012}
{Saripalli}, L., {Subrahmanyan}, R., {Thorat}, K., {et~al.} 2012, \apjs, 199,
  27

\bibitem[{{Scaife} \& {Heald}(2012)}]{scaife2012}
{Scaife}, A.~M.~M. \& {Heald}, G.~H. 2012, \mnras, 423, L30

\bibitem[{{Shimwell} {et~al.}(2016){Shimwell}, {R{\"o}ttgering}, {Best},
  {Williams}, {Dijkema}, {de Gasperin}, {Hardcastle}, {Heald}, {Hoang},
  {Horneffer}, {Intema}, {Mahony}, {Mandal}, {Mechev}, {Morabito}, {Oonk},
  {Rafferty}, {Retana-Montenegro}, {Sabater}, {Tasse}, {van Weeren},
  {Br{\"u}ggen}, {Brunetti}, {Chy{\.z}y}, {Conway}, {Haverkorn}, {Jackson},
  {Jarvis}, {McKean}, {Miley}, {Morganti}, {White}, {Wise}, {van Bemmel},
  {Beck}, {Brienza}, {Bonafede}, {Calistro Rivera}, {Cassano}, {Clarke},
  {Cseh}, {Deller}, {Drabent}, {van Driel}, {Engels}, {Falcke}, {Ferrari},
  {Fr{\"o}hlich}, {Garrett}, {Harwood}, {Heesen}, {Hoeft}, {Horellou},
  {Israel}, {Kapi{\'n}ska}, {Kunert-Bajraszewska}, {McKay}, {Mohan},
  {Orr{\'u}}, {Pizzo}, {Prandoni}, {Schwarz}, {Shulevski}, {Sipior}, {Smith},
  {Sridhar}, {Steinmetz}, {Stroe}, {Varenius}, {van der Werf}, {Zensus}, \&
  {Zwart}}]{shimwell2016}
{Shimwell}, T.~W., {R{\"o}ttgering}, H.~J.~A., {Best}, P.~N., {et~al.} 2016,
  ArXiv e-prints

\bibitem[{{Shulevski} {et~al.}(2017){Shulevski}, {Morganti}, {Harwood},
  {Barthel}, {Jamrozy}, {Brienza}, {Brunetti}, {R{\"o}ttgering}, {Murgia},
  {White}, {Croston}, \& {Br{\"u}ggen}}]{shulevski2017}
{Shulevski}, A., {Morganti}, R., {Harwood}, J.~J., {et~al.} 2017, \aap, 600,
  A65

\bibitem[{{Sirothia} {et~al.}(2009){Sirothia}, {Saikia}, {Ishwara-Chandra}, \&
  {Kantharia}}]{sirothia2009}
{Sirothia}, S.~K., {Saikia}, D.~J., {Ishwara-Chandra}, C.~H., \& {Kantharia},
  N.~G. 2009, \mnras, 392, 1403

\bibitem[{{Slee} {et~al.}(2001){Slee}, {Roy}, {Murgia}, {Andernach}, \&
  {Ehle}}]{slee2001}
{Slee}, O.~B., {Roy}, A.~L., {Murgia}, M., {Andernach}, H., \& {Ehle}, M. 2001,
  \aj, 122, 1172

\bibitem[{{Tasse} {et~al.}(2013){Tasse}, {van der Tol}, {van Zwieten}, {van
  Diepen}, \& {Bhatnagar}}]{tasse2013}
{Tasse}, C., {van der Tol}, S., {van Zwieten}, J., {van Diepen}, G., \&
  {Bhatnagar}, S. 2013, \aap, 553, A105

\bibitem[{{Taylor}(2005)}]{topcat}
{Taylor}, M.~B. 2005, in Astronomical Society of the Pacific Conference Series,
  Vol. 347, Astronomical Data Analysis Software and Systems XIV, ed.
  P.~{Shopbell}, M.~{Britton}, \& R.~{Ebert}, 29

\bibitem[{{Tribble}(1991)}]{tribble1991}
{Tribble}, P.~C. 1991, \mnras, 253, 147

\bibitem[{{Tribble}(1993)}]{tribble1993}
{Tribble}, P.~C. 1993, \mnras, 261, 57

\bibitem[{{Turner} \& {Shabala}(2015)}]{turner2015}
{Turner}, R.~J. \& {Shabala}, S.~S. 2015, \apj, 806, 59

\bibitem[{{van Haarlem} {et~al.}(2013){van Haarlem}, {Wise}, {Gunst}, {Heald},
  {McKean}, {Hessels}, {de Bruyn}, {Nijboer}, {Swinbank}, {Fallows},
  {Brentjens}, {Nelles}, {Beck}, {Falcke}, {Fender}, {H{\"o}randel},
  {Koopmans}, {Mann}, {Miley}, {R{\"o}ttgering}, {Stappers}, {Wijers},
  {Zaroubi}, {van den Akker}, {Alexov}, {Anderson}, {Anderson}, {van Ardenne},
  {Arts}, {Asgekar}, {Avruch}, {Batejat}, {B{\"a}hren}, {Bell}, {Bell}, {van
  Bemmel}, {Bennema}, {Bentum}, {Bernardi}, {Best}, {B{\^i}rzan}, {Bonafede},
  {Boonstra}, {Braun}, {Bregman}, {Breitling}, {van de Brink}, {Broderick},
  {Broekema}, {Brouw}, {Br{\"u}ggen}, {Butcher}, {van Cappellen}, {Ciardi},
  {Coenen}, {Conway}, {Coolen}, {Corstanje}, {Damstra}, {Davies}, {Deller},
  {Dettmar}, {van Diepen}, {Dijkstra}, {Donker}, {Doorduin}, {Dromer}, {Drost},
  {van Duin}, {Eisl{\"o}ffel}, {van Enst}, {Ferrari}, {Frieswijk}, {Gankema},
  {Garrett}, {de Gasperin}, {Gerbers}, {de Geus}, {Grie{\ss}meier}, {Grit},
  {Gruppen}, {Hamaker}, {Hassall}, {Hoeft}, {Holties}, {Horneffer}, {van der
  Horst}, {van Houwelingen}, {Huijgen}, {Iacobelli}, {Intema}, {Jackson},
  {Jelic}, {de Jong}, {Juette}, {Kant}, {Karastergiou}, {Koers}, {Kollen},
  {Kondratiev}, {Kooistra}, {Koopman}, {Koster}, {Kuniyoshi}, {Kramer},
  {Kuper}, {Lambropoulos}, {Law}, {van Leeuwen}, {Lemaitre}, {Loose}, {Maat},
  {Macario}, {Markoff}, {Masters}, {McFadden}, {McKay-Bukowski}, {Meijering},
  {Meulman}, {Mevius}, {Middelberg}, {Millenaar}, {Miller-Jones}, {Mohan},
  {Mol}, {Morawietz}, {Morganti}, {Mulcahy}, {Mulder}, {Munk}, {Nieuwenhuis},
  {van Nieuwpoort}, {Noordam}, {Norden}, {Noutsos}, {Offringa}, {Olofsson},
  {Omar}, {Orr{\'u}}, {Overeem}, {Paas}, {Pandey-Pommier}, {Pandey}, {Pizzo},
  {Polatidis}, {Rafferty}, {Rawlings}, {Reich}, {de Reijer}, {Reitsma},
  {Renting}, {Riemers}, {Rol}, {Romein}, {Roosjen}, {Ruiter}, {Scaife}, {van
  der Schaaf}, {Scheers}, {Schellart}, {Schoenmakers}, {Schoonderbeek},
  {Serylak}, {Shulevski}, {Sluman}, {Smirnov}, {Sobey}, {Spreeuw}, {Steinmetz},
  {Sterks}, {Stiepel}, {Stuurwold}, {Tagger}, {Tang}, {Tasse}, {Thomas},
  {Thoudam}, {Toribio}, {van der Tol}, {Usov}, {van Veelen}, {van der Veen},
  {ter Veen}, {Verbiest}, {Vermeulen}, {Vermaas}, {Vocks}, {Vogt}, {de Vos},
  {van der Wal}, {van Weeren}, {Weggemans}, {Weltevrede}, {White}, {Wijnholds},
  {Wilhelmsson}, {Wucknitz}, {Yatawatta}, {Zarka}, {Zensus}, \& {van
  Zwieten}}]{vanhaarlem2013}
{van Haarlem}, M.~P., {Wise}, M.~W., {Gunst}, A.~W., {et~al.} 2013, \aap, 556,
  A2

\bibitem[{{van Weeren} {et~al.}(2009){van Weeren}, {R{\"o}ttgering},
  {Br{\"u}ggen}, \& {Cohen}}]{vanweeren2009}
{van Weeren}, R.~J., {R{\"o}ttgering}, H.~J.~A., {Br{\"u}ggen}, M., \& {Cohen},
  A. 2009, \aap, 508, 75

\bibitem[{{Wall} \& {Jackson}(1997)}]{wall1997}
{Wall}, J.~V. \& {Jackson}, C.~A. 1997, \mnras, 290, L17

\bibitem[{{Wang} \& {Kaiser}(2008)}]{wang2008}
{Wang}, Y. \& {Kaiser}, C.~R. 2008, \mnras, 388, 677

\bibitem[{{Whittam} {et~al.}(2013){Whittam}, {Riley}, {Green}, {Jarvis},
  {Prandoni}, {Guglielmino}, {Morganti}, {R{\"o}ttgering}, \&
  {Garrett}}]{whittam2013}
{Whittam}, I.~H., {Riley}, J.~M., {Green}, D.~A., {et~al.} 2013, \mnras, 429,
  2080

\bibitem[{{Willott} {et~al.}(2001){Willott}, {Rawlings}, {Blundell}, {Lacy}, \&
  {Eales}}]{willott2001}
{Willott}, C.~J., {Rawlings}, S., {Blundell}, K.~M., {Lacy}, M., \& {Eales},
  S.~A. 2001, \mnras, 322, 536

\bibitem[{{Wilman} {et~al.}(2008){Wilman}, {Miller}, {Jarvis}, {Mauch},
  {Levrier}, {Abdalla}, {Rawlings}, {Kl{\"o}ckner}, {Obreschkow}, {Olteanu}, \&
  {Young}}]{wilman2008}
{Wilman}, R.~J., {Miller}, L., {Jarvis}, M.~J., {et~al.} 2008, \mnras, 388,
  1335

\end{thebibliography}

\end{document}